\documentclass[a4paper,10pt]{revtex4}
\usepackage{mathrsfs}
\usepackage{graphicx}
\usepackage{latexsym}
\usepackage{amsmath}
\usepackage{amssymb}
\usepackage{textcomp}
\usepackage{amsbsy}
\usepackage{graphics}
\usepackage{epstopdf}
\usepackage{color}

\allowdisplaybreaks[4]

\begin{document}

\tolerance=5000

\title{Inflationary magnetogenesis with reheating phase from higher curvature coupling}

\author{Kazuharu Bamba,$^{1}$\,\thanks{bamba@sss.fukushima-u.ac.jp}                  
E.~Elizalde,$^{2}$\,\thanks{elizalde@ieec.uab.es}
S.~D.~Odintsov,$^{2,3,4}$\,\thanks{odintsov@ieec.uab.es}
Tanmoy~Paul$^{5,6}$\,\thanks{pul.tnmy9@gmail.com}} \affiliation{ $^{1)}$ Division of Human Support System, Faculty of Symbiotic Systems Science, 
Fukushima University, Fukushima 960-1296, Japan \\ 
$^{2)}$ Institute of Space Sciences (IEEC-CSIC) C. Can Magrans s/n 08193 Bellaterra (Barcelona), Spain \\
$^{3)}$ ICREA, Passeig Luis Companys 23, 08010 Barcelona, Spain\\
$^{4)}$ Laboratory for Theoretical Cosmology, TUSUR, 634050 Tomsk, Russia.\\
$^{5)}$ Department of Physics, Chandernagore College, Hooghly - 712 136.\\
$^{(6)}$ Department of Theoretical Physics, Indian Association for the Cultivation of Science, 2A $\&$ 2B Raja S.C. Mullick Road, Kolkata - 700 032, India }

\tolerance=5000

\begin{abstract}
We investigate the generation of magnetic fields from inflation, which occurs via breakdown of the conformal 
invariance of the electromagnetic (EM) field, when coupled with the Ricci scalar and the Gauss-Bonnet invariant. For the case of  
instantaneous reheating, the resulting strength of the magnetic field at present is too small and violates the observational 
constraints. However, the problem is solved provided there is a reheating phase with a non-zero e-fold number. During reheating, 
the energy density of the magnetic field is seen to evolve as $(a^3H)^{-2}$ and, after that, as $a^{-4}$ up to the present epoch 
(here $a$ is the scale factor and $H$  the Hubble parameter). It is found that this reheating phase --characterized by a certain e-fold number, 
a constant value of the equation of state parameter, and a given reheating temperature-- renders the magnetogenesis model 
compatible with the observational constraints. The model provides, in turn, a viable way of constraining the reheating equation of 
state parameter, from data analysis of the cosmic microwave background radiation. Moreover we discuss the Schwinger backreaction in the 
present context and determine the necessary constraints on the reheating equation of state parameter.
\end{abstract}

%\pacs{}

\maketitle

\section{Introduction}
Magnetic fields have been observed over the broad range of scales probed so far. 
They have been detected in galaxies, galaxy clusters and even in intergalactic voids 
\cite{Grasso:2000wj,Beck:2000dc,Widrow:2002ud,Kandus:2010nw,Durrer:2013pga,Subramanian:2015lua}. 
Our understanding of the origin of such large scale magnetic fields can be broadly split along two directions. 
The first is associated with an astrophysical origin of the fields, which are later amplified by some dynamo 
mechanism \cite{Kulsrud:2007an,Brandenburg:2004jv,Subramanian:2009fu}. 
The other possibility is that the magnetic fields have a primordial origin, 
i.e., a possible generation of magnetic fields during the 
inflationary epoch \cite{Sharma:2017eps,Sharma:2018kgs,Jain:2012ga,Durrer:2010mq,Kanno:2009ei,Campanelli:2008kh,
Demozzi:2009fu,Bamba:2008ja,Bamba:2008xa,Bamba:2012mi,Bamba:2006ga,Bamba:2003av,Bamba:2004cu,Kobayashi:2019uqs,Bamba:2008my,Giovannini:2017rbc,
Giovannini:2003yn,Lambiase:2004zb,Lambiase:2008zz,Ratra:1991bn,Ade:2015cva,
Chowdhury:2018mhj,Vachaspati:1991nm,Turner:1987bw,Takahashi:2005nd,Agullo:2013tba,Ferreira:2013sqa,Atmjeet:2014cxa}, or in alternative scenarios, as in 
 bouncing cosmology \cite{Frion:2020bxc,Chowdhury:2016aet,Chowdhury:2018blx,Koley:2016jdw,Qian:2016lbf,Membiela:2013cea}. A confirmation of the announced detection of magnetic fields in the large voids might in principle enforce their primordial origin, in front of the other possibility.

Among all the proposals discussed so far, the primordial origin of magnetic fields during inflation has earned  a lot of attention, primarily because the inflationary paradigm is able to solve the horizon and flatness problems, and to generate an almost scale invariant power spectrum, which is perfectly consistent with the observational data 
\cite{guth,Linde:2005ht,Langlois:2004de,Riotto:2002yw,Baumann:2009ds,Bamba:2015uma}. However, the corresponding inflationary magnetogenesis 
is riddled with severe difficulties: the most crucial one is how to generate a value of the magnetic strength that is high enough to be compatible with present day observations at the galactic scale. In the standard Maxwell theory, the electromagnetic (EM) field is endowed with a conformal symmetry, so that the electromagnetic field energy density decays as $a^{-4}$ with the universe expansion. Such behavior leads to a very feeble value of the magnetic strength at present, which is fully unable to account for the observational results. This is a clear indication that, in the context of 
inflationary magnetogenesis, the conformal invariance of the electromagnetic field should be broken at an early stage, which in turn would allow 
gauge field production from the quantum vacuum state, thus preventing the electromagnetic field energy from decaying as fast as $a^{-4}$. 
Several models have been already proposed in the literature to break the conformal invariance of the electromagnetic action, as those including a 
non-trivial coupling between a scalar field (generally considered as the inflaton) and the gauge field 
\cite{Sharma:2017eps,Sharma:2018kgs,Jain:2012ga,Durrer:2010mq,Kanno:2009ei,Campanelli:2008kh,
Demozzi:2009fu,Bamba:2008ja,Bamba:2008xa,Bamba:2012mi,Bamba:2006ga,Bamba:2003av,Bamba:2004cu,Kobayashi:2019uqs,Bamba:2008my,Giovannini:2017rbc,
Giovannini:2003yn,Lambiase:2004zb,Lambiase:2008zz,Ratra:1991bn,Ade:2015cva,
Chowdhury:2018mhj,Vachaspati:1991nm,Turner:1987bw,Takahashi:2005nd,Agullo:2013tba,Ferreira:2013sqa,Atmjeet:2014cxa,Caprini:2014mja,
Kobayashi:2014sga,Atmjeet:2013yta,Fujita:2015iga,Campanelli:2015jfa,Tasinato:2014fia}. Alternative scenarios to break the conformal invariance 
have been proposed in the shelter of non-linear electrodynamics \cite{Campanelli:2007cg} or 3-form fields \cite{Urban:2013aka}. 
Other important issues that affect inflationary 
magnetogenesis models are the backreaction and the strong coupling problems \cite{Sharma:2017eps,Demozzi:2009fu,Ferreira:2013sqa,
Kobayashi:2014sga,Markkanen:2017kmy,Tasinato:2014fia}. 
The backreaction issue appears when the 
strength of the electromagnetic field exceeds the background energy density, which in turn may spoil the inflationary set-up as well, as it
suppresses the production of magnetic fields. On the other hand, the strong coupling problem occurs if the effective electric charge becomes 
high during inflation, rendering the perturbative calculation of the EM field unreliable.

Apart from the inflationary magnetogenesis set-up, one should also add some proposals for magnetic field generation from  entirely different perspectives, in 
particular, from the bouncing scenario 
\cite{Frion:2020bxc,Chowdhury:2016aet,Chowdhury:2018blx,Koley:2016jdw,Qian:2016lbf,Membiela:2013cea}. 
Similar to the inflationary theory, bounce cosmology is also able to 
generate a nearly scale invariant power spectrum, thus becoming compatible with the observational data available. However, most of the bouncing models 
are plagued with certain difficulties in regard to the cosmological background and the evolution of perturbations; in particular, the BKL instability associated 
with the anisotropic problem, the violation of the energy conditions at the bounce point, the instability of scalar and tensor perturbations, 
etc. \cite{Brandenberger:2012zb,Brandenberger:2016vhg,Battefeld:2014uga,Novello:2008ra,Cai:2014bea,Nojiri:2019lqw,Odintsov:2015ynk}. Here it 
should be mentioned that these  severe 
problems can be solved, to some extent, in various modified theories of gravity \cite{Battefeld:2014uga,Cai:2008qw,Cai:2016thi,
Elizalde:2019tee,Elizalde:2020zcb,Navo:2020eqt,Bamba:2014mya,Odintsov:2020zct,Banerjee:2020uil}.

In this paper, we propose an inflationary magnetogenesis model, in which the electromagnetic field couples to the background spacetime curvature, specifically 
with the Ricci scalar and the Gauss-Bonnet invariant; i.e we include in our model a higher curvature coupling of the EM field. 
Such  coupling  breaks the conformal invariance of the electromagnetic action and allows the production of photons from the Bunch-Davies vacuum. 
Moreover, being this a higher-order operator, the coupling is suppressed at the Planck scale and, thus, the model becomes free 
from the strong coupling problem. In regard to the background spacetime, we consider the scalar-Einstein-Gauss-Bonnet gravity theory, which 
is known to provide viable inflationary models (consistent with the latest Planck results), for suitable choices of the Gauss-Bonnet 
coupling function and the scalar field potential \cite{Li:2007jm,Odintsov:2018nch,Carter:2005fu,Nojiri:2019dwl,Elizalde:2010jx,
Makarenko:2016jsy,delaCruzDombriz:2011wn,Bamba:2007ef,Chakraborty:2018scm,
Kanti:2015pda,Kanti:2015dra,Odintsov:2018zhw,Saridakis:2017rdo,Cognola:2006eg}. In such scenario, we try to explore the dynamics of the electric 
and magnetic fields along with the cosmological 
expansion of the universe, starting from the inflationary stage. During the first steps of the cosmic expansion, the universe 
enters  a reheating phase, after the end of inflation; and depending on the reheating mechanism, we consider two different scenarios: 
(1) instantaneous reheating 
at the end of inflation and (2) a Kamionkowski like reheating model with non-zero e-fold number, in which case the reheating phase is parametrized 
by a constant effective equation of state (EoS) parameter ($\omega_\mathrm{eff}$) \cite{Dai:2014jja} (for recent results on the reheating phase, see 
\cite{Albrecht:1982mp,Ellis:2015pla,Ueno:2016dim,Eshaghi:2016kne,Maity:2018qhi,Haque:2020zco,DiMarco:2017zek,Drewes:2017fmn,DiMarco:2018bnw}). 
These two scenarios make qualitative differences in the evolution 
of electric and magnetic fields. In particular, the presence of the reheating phase with a non-zero e-fold number 
enhances the strength of the magnetic field, in comparison 
with the instantaneous reheating case, and this is reflected in the present amplitude of the magnetic field, which is found to differ in the two cases.
We should mention that  magnetogenesis models with curvature couplings have been proposed earlier, however in quite different contexts 
\cite{Turner:1987bw,Kushwaha:2020nfa,Guo:2015awg}. Note that, in our present analysis, we include the higher 
curvature Gauss-Bonnet coupling in the set-up and also discuss the effect of the reheating phase in the production of the magnetic field, 
which makes the present scenario essentially different from earlier ones. Moreover at the time of preparing our manuscript, some authors 
investigated the effect of reheating phase in an inflationary magnetogenesis model \cite{Haque:2020bip}, 
however without introducing any higher curvature coupling in the model.

The paper is organized as follows: after describing the model in Sec.~\ref{sec_model}, we give the general expressions for the EM power spectra in the 
present context in Sec.~\ref{sec_general expression}. The solution for the vector potential and the electromagnetic energy density during inflation 
are presented in Sec.~\ref{sec_solution_inflation}. The corresponding calculations during the reheating phase are carried out in 
Sec.~\ref{sec_instantaneous reheating} and Sec.~\ref{sec_kamionkowski}, which  correspond to the cases of  instantaneous reheating and a 
Kamionkowski like reheating model, respectively. The paper ends with some conclusions.

\section{The model}\label{sec_model}
Consider the following action,
\begin{eqnarray}
 S = S_{grav} + S_{em}^{(can)} + S_{CB},
 \label{action0}
\end{eqnarray}
where $S_{grav}$ symbolizes the action for the underlying gravity theory which we consider as a scalar coupled Einstein-Gauss-Bonnet theory. 
In the most general setting, the action for the scalar coupled Einstein-Gauss-Bonnet gravity consists 
of four terms — the Ricci scalar, the Gauss-Bonnet invariant coupled to an arbitrary function of the scalar field, the
kinetic term of the scalar field, and a self-interaction term for the scalar field, such that 
\begin{eqnarray}
 S_{grav} = \int d^4x\sqrt{-g}\bigg[\frac{R}{2\kappa^2} - \frac{1}{2}\partial_{\mu}\Phi\partial_{\nu}\Phi - V(\Phi) - \xi(\Phi)\mathcal{G}\bigg]
 \label{action part 1}
\end{eqnarray}
where $R$ is the Ricci scalar, $\kappa^2 = 8\pi G$ ($G$ is Newton's constant), $\Phi$ is the scalar field generally known as 
inflaton field embedded within the potential $V(\Phi)$ and 
$\mathcal{G} = R^2 - 4R_{\mu\nu}R^{\mu\nu} + R_{\mu\nu\alpha\beta}R^{\mu\nu\alpha\beta}$ is the Gauss-Bonnet invariant. 
The presence of the coupling function between the scalar field and the Gauss-Bonnet term, symbolized by $\xi(\Phi)$, ensures 
the non-topological character of the Gauss-Bonnet term in the above action. The second term of the action (\ref{action0}), i.e $S_{em}^{(can)}$, 
denotes the standard electromagnetic field action and given by, 
\begin{eqnarray}
 S_{em}^{(can)} = \int d^4x\sqrt{-g}\big[-\frac{1}{4}F_{\mu\nu}F^{\mu\nu}\big]
 \label{action part 2}
\end{eqnarray}
where $F_{\mu\nu} = \partial_{\mu}A_{\nu} - \partial_{\nu}A_{\mu}$ is the electromagnetic field tensor of the vector field $A_{\mu}$. Finally 
$S_{CB}$ refers to the conformal symmetry breaking part  and, in the present context, we consider a non-minimal curvature coupling 
which breaks the conformal invariance of the electromagnetic field. In particular $S_{CB}$ is given by,
\begin{eqnarray}
 S_{CB} = \int d^4x\sqrt{-g} \big[f(R,\mathcal{G})F_{\mu\nu}F^{\mu\nu}\big]
 \label{actuon part 3}
\end{eqnarray}
where $f(R,\mathcal{G})$ is an arbitrary analytic function of the Ricci scalar and the GB invariant at the moment and denotes the curvature coupling of the 
electromagnetic field. $f(R,\mathcal{G})$, in the present context, is considered to be a polynomial function of the Ricci scalar and the GB 
invariant, in particular, 
\begin{eqnarray}
 f(R,\mathcal{G}) = \kappa^{2q}\big(R^q + \mathcal{G}^{q/2}\big),
 \label{form of f 1}
\end{eqnarray}
where $q$ is the model parameter. The form of $f(R,\mathcal{G})$ clearly indicates that, as the curvature is significant in the early universe, the curvature 
coupling introduces then a non-trivial correction to the electromagnetic action; however, at late times (in particular after the end of inflation, as we will 
show at a later stage), $S_{CB}$ will not contribute and the electromagnetic field will then behave according to the standard Maxwell's equations. This is not the case in most of the earlier magnetogenesis models, where a  non-minimal coupling between scalar and electromagnetic field is considered in the action in order 
to break the conformal invariance. Here 
we would like to stress that the conformal breaking term $f(R,\mathcal{G})$ is suppressed by $\kappa^{2q}$ and, thus, the present model is free from the 
strong coupling problem for $q \sim \mathcal{O}(1)$, what is a remarkable feature of our proposal. 
Moreover,  we will show later that the electromagnetic field has a negligible backreaction on the background 
inflationary FRW spacetime and, thus, the backreaction problem in the present magnetogenesis scenario will be also resolved naturally.

In regard to the background spacetime evolution, it is worthwhile to mention that the scalar-Einstein-GB theory of gravity leads to an inflationary scenario 
which is indeed stable with respect to scalar and tensor perturbations of the FRW metric, for suitable choices of $V(\Phi)$ and $\xi(\Phi)$. 
For example, for  quadratic choices of $V(\phi)$ 
and $\xi(\Phi)$, one can show that $S_{grav}$ provides an acceleration phase of the early universe, which has a graceful exit for numerically interpolated 
forms of $V(\Phi)$ and $\xi(\Phi)$, starting from the quadratic function of the scalar field \cite{Chakraborty:2018scm}. 
The stability of scalar and tensor perturbations in the context of the
Gauss-Bonnet gravity theory are ensured due to the presence of the scalar field potential $V(\Phi)$ in the gravitational action. Furthermore, 
the speed of the tensor perturbations ($c_T^2$), in general, is not unity in the scalar coupled Einstein-GB theory and the deviation 
of $c_T^2$ from unity is proportional to the GB coupling function considered in the model. The result $c_T^2 \neq 1$ precludes that the gravitational waves 
propagate with a different speed, compared to the speed of light which is unity in natural units, and thus is not in agreement with the event 
GW170817. However, there exists a certain class of GB coupling function for which the gravitational wave propagates with $c_T^2 = 1$ leading to 
the compatibility of the GB model with GW170817. The inflationary phenomenology and its viability with the latest Planck 2018 results in such Gauss-Bonnet 
theory that is compatible with GW170817 have been recently discussed in \cite{Odintsov:2019clh,Odintsov:2020sqy,Odintsov:2020mkz}.

With the action (\ref{action0}), particularly with the conformal breaking term $S_{CB}$, we aim to generate a sufficiently strong magnetic field 
in the present epoch. The variation of action (\ref{action0}) with respect to the gauge field leads to the following equation of motion for $A_{\mu}$,
\begin{eqnarray}
 \partial_{\alpha}\bigg[\sqrt{-g}h^2(R,G)~g^{\mu\alpha}g^{\nu\beta}F_{\mu\nu}\bigg] = 0
 \label{eom}
\end{eqnarray}
with $h^2(R,\mathcal{G}) = 1 - 4f(R,\mathcal{G})$. The spatially flat FRW metric ansatz will fulfill our purpose i.e we take,
\begin{eqnarray}
 ds^2 = a^2(\eta)\big[-d\eta^2 + d\vec{x}^2\big]
 \label{FRW metric ansatz}
\end{eqnarray}
where $\eta$ is known as conformal time and $a(\eta)$ is the scale factor. Owing to the FRW metric ansatz, the temporal and spatial component 
of Eq.(\ref{eom}) reduce to
\begin{eqnarray}
 \partial^{i}\big[\partial_iA_0 - \partial_0A_i\big] = 0
 \label{temporal eom}
 \end{eqnarray}
 and
 \begin{eqnarray}
 \delta^{ij}~\partial_0\big[h^2(R,G)F_{0j}\big] - \delta^{ij}~\partial^{l}\big[h^2(R,G)F_{lj}\big]&=&0
 \label{eom1}
\end{eqnarray}
respectively. In the Coulomb gauge ($A_0 = 0$), Eq.(\ref{temporal eom}) leads to the condition $\partial_{i}A_{i} = 0$ which further simplify 
Eq.(\ref{eom1}) as follows,
\begin{eqnarray}
 A_i''(\eta,\vec{x}) + 2\frac{h'(R,G)}{h(R,G)}A_i' - \partial_l\partial^{l}A_i = 0~~.
 \label{eom2}
\end{eqnarray}
As mentioned earlier the gravitational action considered in the present context (see Eq.(\ref{action part 1})) provides a viable inflationary scenario for 
suitable forms of $V(\Phi)$ and $\xi(\Phi)$. Thus, we consider a quasi de-Sitter inflationary scenario as the background spacetime, where the scale factor 
is given by
\begin{eqnarray}
 a(\eta) = \bigg(\frac{-\eta}{\eta_0}\bigg)^{\beta + 1}~~~~~~~~\mathrm{with}~~~~~~~~~\beta = -2-\epsilon = -3 + \frac{\mathcal{H}'}{\mathcal{H}^2}.
 \label{scale factor}
\end{eqnarray}
Here a prime denotes differentiation with respect to $\frac{d}{d\eta}$, 
$\mathcal{H} = \frac{a'}{a}$ is conformal the Hubble parameter and $\epsilon$ is the slow roll 
parameter having the expression $\epsilon = -\frac{\mathcal{H}'}{\mathcal{H}^2} + 1$. The scale factor of Eq.(\ref{scale factor}) immediately leads to 
the Hubble parameter, Ricci scalar and the Gauss-Bonnet invariant, as follows
\begin{eqnarray}
 \mathcal{H} = \frac{\beta + 1}{\eta}
 \label{Hubble parameter}
\end{eqnarray}
and
\begin{eqnarray}
 R&=&\frac{6}{a^2}\big(\mathcal{H}' + \mathcal{H}^2\big) = \frac{6\beta(\beta + 1)}{\eta_0^2}\bigg(\frac{-\eta}{\eta_0}\bigg)^{2\epsilon}\nonumber\\
 \mathcal{G}&=&\frac{24}{a^4}\mathcal{H}^2\mathcal{H}' = -\frac{24(\beta + 1)^3}{\eta_0^4}\bigg(\frac{-\eta}{\eta_0}\bigg)^{4\epsilon},
 \label{R and G}
\end{eqnarray}
respectively. The conformal Hubble parameter is related to the cosmic Hubble parameter by 
$H = \frac{1}{a}\mathcal{H}$ and thus we get $H = \frac{1}{\eta_0}\big(-\eta\big)^{\epsilon}$ or in terms of e-folding number
\begin{eqnarray}
 H = \frac{1}{\eta_0}\exp{\bigg(-\frac{\epsilon N}{1 + \epsilon}\bigg)},
 \label{cosmic Hubble parameter}
\end{eqnarray}
where the e-folding number (up to time $\eta$) is defined as $N = \int^{\eta}aH~d\eta$, i.e the beginning of inflation is designated by $N = 0$ and we 
consider it to happen when the CMB scale mode crosses the horizon. Eq.(\ref{cosmic Hubble parameter}) shows that $\frac{1}{\eta_0}$ 
specifies the cosmic Hubble parameter at the starting of inflation (we will denote it by $H_0$, i.e $\eta_0^{-1} = H_0$, in the subsequent calculation). 
Consequently, with the expressions of $R$ and $\mathcal{G}$, we determine 
the curvature coupling function of the electromagnetic field, i.e $f(R,\mathcal{G})$, from Eq.(\ref{form of f 1}) and is given by,
\begin{eqnarray}
 f(R,\mathcal{G}) = \kappa^{2q}\bigg\{\frac{\big[6\beta(\beta+1)\big]^q + \big[-24(\beta+1)^3\big]^{q/2}}{\eta_0^{2q}}\bigg\}\bigg(\frac{-\eta}{\eta_0}\bigg)^{2\epsilon q}
 \label{form of f 2}
\end{eqnarray}
At this stage we would like to mention that the time dependence of $f(R,\mathcal{G})$ is the sole reason to spoil the conformal 
symmetry of the electromagnetic field. However the above equation indicates that $f(R,\mathcal{G})$ becomes constant under 
the condition $\epsilon = 0$ and thus leads to a conformal invariant electromagnetic action. 
Actually, for $\epsilon = 0$ (i.e a de-Sitter background spacetime), the Ricci scalar and the GB invariant become constant and henceforth 
$f(R,\mathcal{G})$ will be, too. Thereby the present model where the electromagnetic field is coupled with the background spacetime curvature, requires 
$\epsilon \neq 0$ in order to break the conformal invariance of the gauge field. Keeping this in mind, we will consider the background 
inflationary scenario as a quasi de-Sitter evolution, in which case $\epsilon \neq 0$ and $\epsilon < 1$, in the subsequent calculation.

\section{Energy density and power spectra for electric and magnetic field}\label{sec_general expression}
In this section, we aim to calculate the power spectra for both the electric and magnetic fields and in this regard, it may be mentioned that 
both fields are intrinsically frame dependent. Here we consider the comoving observer for which the four velocity 
components are given by $u^{\mu} = \big(1/a(\eta), 0, 0, 0\big)$ and thus the proper time of a comoving observer is defined as 
$dt = a(\eta)d\eta$. The computation of the power spectrum requires two ingredients; first we need to know the energy density separately 
for electric, magnetic fields and, secondly, the vacuum state associated with the electromagnetic field in the background inflationary evolution. 
Thereby, from the action (\ref{action0}), we determine the energy-momentum tensor associated with the electromagnetic field 
\begin{eqnarray}
 T_{\alpha\beta}&=&-\frac{2}{\sqrt{-g}}~\frac{\delta}{\delta g^{\alpha\beta}}\bigg[-\frac{1}{4}\sqrt{-g}~\big(1 - 4f(R,\mathcal{G})\big)F_{\mu\nu}F^{\mu\nu}\bigg]\nonumber\\
 &=&-\frac{1}{4}\bigg\{g_{\alpha\beta}\big(1 - 4f(R,\mathcal{G})\big)F_{\mu\nu}F^{\mu\nu} - 4\big(1 - 4f(R,\mathcal{G})\big)g^{\mu\nu}F_{\mu\alpha}F_{\nu\beta} 
 + 8F_{\mu\nu}F^{\mu\nu}\frac{\delta f(R,\mathcal{G})}{\delta g^{\alpha\beta}}\bigg\}~~.
 \label{em tensor}
\end{eqnarray}
The energy density of the electromagnetic field in the background FRW spacetime is defined as $T^0_0 = -\frac{1}{a^2}T_{00}$ and thus 
the above expression of $T_{\alpha\beta}$ immediately leads to the following form of $T^0_0$:
\begin{eqnarray}
 T^0_0 = -\frac{1}{2a^4}\big(A_i'\big)^2\bigg\{\big(1 - 4f(R,\mathcal{G})\big) + \frac{8}{a^2}\frac{\delta f}{\delta g^{00}}\bigg\} 
 - \frac{1}{4a^4}F_{ij}F_{ij}\bigg\{\big(1 - 4f(R,\mathcal{G})\big) - \frac{8}{a^2}\frac{\delta f}{\delta g^{00}}\bigg\}
 \label{energy density 1}
\end{eqnarray}
where we use the Coulomb gauge condition and also the result $F_{\mu\nu}F^{\mu\nu} = \frac{1}{a^4}\big(-2(A_i')^2 + F_{ij}F_{ij}\big)$ 
holds in the spatially flat FRW metric. Furthermore, the term $\frac{\delta f(R,\mathcal{G})}{\delta g^{00}}$ present in the above expression is 
determined as  
\begin{eqnarray}
 \frac{\delta f(R,\mathcal{G})}{\delta g^{00}} 
 = \frac{\partial f}{\partial R}\frac{\delta R}{\delta g^{00}} + \frac{\partial f}{\partial \mathcal{G}}\frac{\delta \mathcal{G}}{\delta g^{00}} 
 = -3q\kappa^{2q}\mathcal{H}'\bigg[R^{q-1} + \frac{2\mathcal{H}^2}{a^2}\mathcal{G}^{\frac{q}{2} - 1}\bigg]~~.
 \nonumber
\end{eqnarray}
Thereby, the final expression of the electromagnetic energy density in the present context is 
\begin{eqnarray}
 T^0_0 = -\frac{1}{2a^4}\big(A_i'\big)^2 P(\eta) - \frac{1}{4a^4}F_{ij}F_{ij} Q(\eta),
 \label{energy density 2}
\end{eqnarray}
where $P(\eta)$ and $Q(\eta)$ have the following form
\begin{eqnarray}
 P(\eta)&=&1 - 4\kappa^{2q}\big(R^{q} + \mathcal{G}^{q/2}\big) 
 - \frac{24q\kappa^{2q}\mathcal{H}'}{a^2}\bigg(R^{q-1} + \frac{2\mathcal{H}^2}{a^2}\mathcal{G}^{\frac{q}{2} - 1}\bigg)\nonumber\\
 Q(\eta)&=&1 - 4\kappa^{2q}\big(R^{q} + \mathcal{G}^{q/2}\big) 
 + \frac{24q\kappa^{2q}\mathcal{H}'}{a^2}\bigg(R^{q-1} + \frac{2\mathcal{H}^2}{a^2}\mathcal{G}^{\frac{q}{2} - 1}\bigg),
 \label{P and Q}
\end{eqnarray}
with $R = R(\eta)$ and $\mathcal{G} = \mathcal{G}(\eta)$ as shown in Eq.(\ref{R and G}). Thus, using Eq.(\ref{R and G}), 
one can further determine $P(\eta)$ and $Q(\eta)$ as functions of 
the conformal time 
\begin{eqnarray}
 P(\eta) = 1 - \frac{4B}{\eta_0^{2q}}\bigg(\frac{-\eta}{\eta_0}\bigg)^{2\epsilon q} 
 - \frac{24q\kappa^{2q}~\mathcal{H}'}{a^2}\bigg\{\frac{\big[6\beta(\beta+1)\big]^{q-1}}{\eta_0^{2q-2}}
 \bigg(\frac{-\eta}{\eta_0}\bigg)^{2\epsilon q - 2\epsilon} + \frac{2\mathcal{H}^2\big[-24(\beta+1)^3\big]^{q/2-1}}{\eta_0^{2q-4}}
 \bigg(\frac{-\eta}{\eta_0}\bigg)^{2\epsilon q - 4\epsilon}\bigg\}
 \label{P}
\end{eqnarray}
and
\begin{eqnarray}
Q(\eta) = 1 - \frac{4B}{\eta_0^{2q}}\bigg(\frac{-\eta}{\eta_0}\bigg)^{2\epsilon q} 
 + \frac{24q\kappa^{2q}~\mathcal{H}'}{a^2}\bigg\{\frac{\big[6\beta(\beta+1)\big]^{q-1}}{\eta_0^{2q-2}}
 \bigg(\frac{-\eta}{\eta_0}\bigg)^{2\epsilon q - 2\epsilon} + \frac{2\mathcal{H}^2\big[-24(\beta+1)^3\big]^{q/2-1}}{\eta_0^{2q-4}}
 \bigg(\frac{-\eta}{\eta_0}\bigg)^{2\epsilon q - 4\epsilon}\bigg\},
 \label{Q}
\end{eqnarray}
respectively, with $B = \kappa^{2q}\bigg[\big[6\beta(\beta + 1)\big]^q + \big[-24(\beta + 1)^3\big]^{q/2}\bigg]$ and $\mathcal{H} = \mathcal{H}(\eta)$ 
as given in Eq.(\ref{Hubble parameter}). It is clear that the functions $P(\eta)$ and $Q(\eta)$ deviate from unity and become non-trivial solely 
due to the presence of the conformal breaking term $f(R,\mathcal{G})$ in the electromagnetic action. 
Having determined $T_0^0$, we are now in the position to separate the energy density of the electric ($\vec{E}$) 
and magnetic ($\vec{B}$) fields, respectively. The first term in Eq.(\ref{energy density 2}) is obviously the energy density of the electric field, 
while the other one, depending only on the spatial derivatives of the vector potential, contributes to the magnetic field. Hence, the expectation value 
of the electric field energy density over the vacuum state $|0\rangle$ can be written down as
\begin{eqnarray}
 \rho(\vec{E}) = -\frac{P(\eta)}{2a^4}~\langle 0|(A_i')^2|0 \rangle~~.
 \label{expectation energy density electric1}
\end{eqnarray}
Similarly, the expectation energy density of the magnetic field over $|0\rangle$ reads
\begin{eqnarray}
 \rho(\vec{B}) = -\frac{Q(\eta)}{4a^4}~\langle 0|F_{ij}F_{ij}| 0 \rangle~~.
 \label{expectation energy density magnetic1}
\end{eqnarray}
Here $|0\rangle$ is the distant past vacuum state for the electromagnetic field and later we will show that the Bunch-Davies state 
can act as a suitable infinite past vacuum for the gauge field. To evaluate $\rho(\vec{E})$ and $\rho(\vec{B})$ explicitly, we need to quantize 
the gauge field (over the inflationary background) by promoting $A_i(\eta,\vec{x})$ to a hermitian operator $\hat{A}_i(\eta,\vec{x})$ and expanding 
it in a Fourier basis, as
\begin{eqnarray}
 \hat{A}_i(\eta,\vec{x}) = \int \frac{d\vec{k}}{(2\pi)^3}\sum_{r=1,2}\epsilon_{ri}~\bigg[\hat{b}_r(\vec{k})A_{r}(k,\eta)e^{i\vec{k}.\vec{x}} 
 + \hat{b}_r^{+}(\vec{k})A_{r}^{*}(k,\eta)e^{-i\vec{k}.\vec{x}}\bigg],
 \label{mode decomposition}
\end{eqnarray}
where $\vec{k}$ is the Fourier mode momentum (or equivalently the electromagnetic wave vector), 
$r$ is the polarization index and runs from $r = 1,2$ with $\epsilon_{ri}$ being the two polarization vectors. Here we consider the polarization vectors 
in the standard linear polarization basis, in which case $\epsilon_{1i} = (1, 0, 0)$ and $\epsilon_{2i} = (0, 1, 0)$. Clearly in such 
polarization basis, the Coulomb gauge condition, characterized by $\partial_iA^{i} = 0$, further leads to the condition $k^{i}\epsilon_{1i} = 
k^{i}\epsilon_{2i} = 0$, which states that the propagation direction of the electromagnetic wave (or the propagation direction of photon in the 
quantized language) is perpendicular to the plane spanned by the polarization vectors. 
Moreover, $\hat{b}_r(\vec{k})$ and 
$\hat{b}_r^{+}(\vec{k})$ are the annihilation and creation operators defined on the distant past vacuum state $|0\rangle$, i.e 
the relation $\hat{b}_r(\vec{k})|0\rangle = 0$ holds for all $\vec{k}$. These creation and annihilation operators follow the quantization rule, as
\begin{eqnarray}
 \big[\hat{b}_p(\vec{k}),\hat{b}_r^{+}(\vec{k}')\big] = \delta_{pr}\delta\big(\vec{k} - \vec{k}'\big).
 \label{commutation relation}
\end{eqnarray}
With the mode decomposition of $A_i(\eta,\vec{x})$ expressed in Eq.(\ref{mode decomposition}) along with the above commutation relation, the expectation 
energy densities of electric and magnetic fields given in Eqs.(\ref{expectation energy density electric1}) and (\ref{expectation energy density magnetic1}), 
respectively reduce to the following expressions
\begin{eqnarray}
 \rho(\vec{E}) = P(\eta) \sum_{r=1,2}\int \frac{dk}{2\pi^2}~\frac{k^2}{a^4}\big|A_r'(k,\eta)\big|^2\nonumber\\
 \rho(\vec{B}) = Q(\eta) \sum_{r=1,2}\int \frac{dk}{2\pi^2}~\frac{k^4}{a^4}\big|A_r(k,\eta)\big|^2~~.
 \label{expectation energy density}
\end{eqnarray}
Consequently, the power spectra (defined as the energy density associated to a logarithmic interval of $k$) 
of the electric and magnetic fields follow
\begin{eqnarray}
 \frac{\partial \rho(\vec{E})}{\partial \ln{k}} = P(\eta) \sum_{r=1,2} \frac{k}{2\pi^2}~\frac{k^2}{a^4}\big|A_r'(k,\eta)\big|^2~~~~~~~~~~~~,~~~~~~~~~~~
 \frac{\partial \rho(\vec{B})}{\partial \ln{k}} = Q(\eta) \sum_{r=1,2} \frac{k}{2\pi^2}~\frac{k^4}{a^4}\big|A_r(k,\eta)\big|^2,
 \label{power spectra}
\end{eqnarray}
where $P(\eta)$ and $Q(\eta)$ are shown in Eqs.(\ref{P}) and (\ref{Q}), respectively. The appearance of the non-trivial functions $P(\eta)$ and $Q(\eta)$ 
make the electric and magnetic power spectra in the present context different in comparison to those of the standard electrodynamic case and, moreover, 
as mentioned earlier, $P(\eta)$ and $Q(\eta)$ deviate from unity solely due to the effect of the conformal breaking coupling $f(R,\mathcal{G})$. 
The $k$ dependence in the electric and magnetic power spectra are seemingly different from Eq.(\ref{power spectra}), in particular, in the electric power 
spectrum the $k$ dependence comes through $k^3$ and the time derivative of the mode function while for the magnetic case it comes through $k^5$ and the 
mode function itself. Thus, it is very much likely that, when the electric spectrum becomes scale invariant, the magnetic spectrum is not so, and vice-versa. 
In order to reveal the explicit $k$-dependence of the power spectra, we need to solve the mode function, which is the subject of the next section. 
Further, the time dependence of the power spectra is also important to understand, in order to investigate the backreaction problem, which also requires the solution 
of the electromagnetic mode function.

\section{Solving for the mode function and scale dependence of the power spectra during inflation}\label{sec_solution_inflation}
\begin{figure}[!h]
\begin{center}
 \centering
 \includegraphics[width=3.5in,height=2.0in]{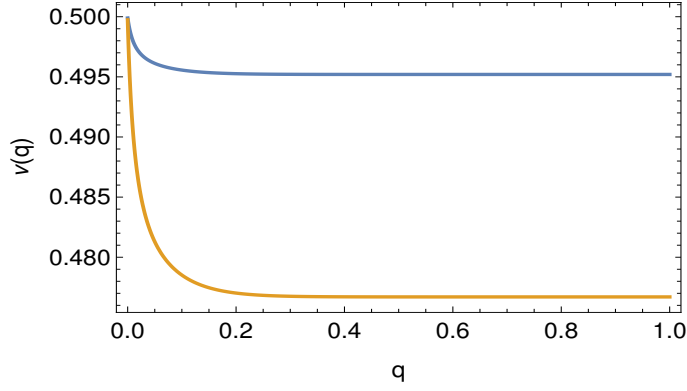}
 \caption{$\nu(q)$ vs $q$: In the $Blue~curve$, $H_0 = 10^{-5}M_{Pl}$, $\epsilon = 0.01$ and in the $Yellow~Curve$, $H = 10^{-5}M_{Pl}$, $\epsilon = 0.1$. 
 Both the curves start from $\nu = 0.5$ at $q = 0$, which is also expected from the expression of $\nu(q)$.} 
 \label{plot1}
\end{center}
\end{figure}

In this section we will determine the solution of the mode function and for this purpose we need the evolution equation for the vector potential 
in terms of the conformal time, which has been already written down in Eq.(\ref{eom2}). More explicitly, we need to recast Eq.(\ref{eom2}) 
in Fourier space, which leads to the evolution equation of the mode function as
\begin{eqnarray}
 A_r''(k,\eta) + \frac{2h'}{h}A_r'(k,\eta) + k^2A_r(k,\eta) = 0,
 \label{FT eom1}
\end{eqnarray}
where $h^2(R,\mathcal{G}) = 1 - 4f(R,\mathcal{G})$ and $f(R,\mathcal{G})$ is expressed in Eq.(\ref{form of f 2}). Introducing 
Mukhanov-Sasaki like variable for the electromagnetic field as $\tilde{A}_r(k,\eta) = h(\eta)A_r(k,\eta)$, the above equation transforms to
\begin{eqnarray}
 \tilde{A}_r''(k,\eta) + \bigg(k^2 - \frac{h''}{h}\bigg)\tilde{A}_r(k,\eta) = 0~~~.
 \label{FT eom2}
\end{eqnarray}
The factor $\frac{h''}{h}$ entirely depends on the background spacetime evolution and, thus, the last term in the left hand side of Eq.(\ref{FT eom2}) 
depicts how the electromagnetic perturbation 
couples with the background spacetime curvature. Using Eq.(\ref{form of f 2}), the $h(R,\mathcal{G})$ is determined as 
$h(R,\mathcal{G}) = \big(1 - 4f(R,\mathcal{G})\big)^{1/2} = \bigg[1 - \frac{4B}{\eta_0^{2q}}\bigg(\frac{-\eta}{\eta_0}\bigg)^{2\epsilon q}\bigg]^{1/2}$ 
with recall, $B = \kappa^{2q}\bigg[\big[6\beta(\beta + 1)\big]^q + \big[-24(\beta + 1)^3\big]^{q/2}\bigg]$. At this stage, it deserves mentioning that 
the conformal breaking (CB) coupling $f(R,\mathcal{G})$ is suppressed by the Planck mass over the exponent $2q$ and, thus, a higher value of $q$ leads to a 
larger suppression of $f(R,\mathcal{G})$ by the Planck scale. Hence, we consider $0 < q < 1$ in the present work (the negative values of $q$ will 
lead to a divergence at $R \rightarrow 0$ and, thus, we exclude the case $q < 0$). Such consideration of $q$ along 
with $\epsilon < 1$ leads to the condition $f(R,\mathcal{G}) < 1$ for a wide range of conformal time, starting deeply from the sub-Hubble regime; 
as an example for $q = 0.5$ and $\epsilon = 0.01$, 
then $f(R,\mathcal{G})$ becomes less than unity in the regime $-k\eta < 10^{332}$ which is in the deep sub-Hubble radius. This safely allows 
to expand $h(R,\mathcal{G}) = \big[1 - 4f(R,\mathcal{G})\big]^{1/2}$ as a binomial expansion, which leads to the following expression 
\begin{eqnarray}
 h(R,\mathcal{G}) = 1 - \frac{2B}{\eta_0^{2q}}\bigg(\frac{-\eta}{\eta_0}\bigg)^{2\epsilon q},
 \nonumber
\end{eqnarray}
where we retain upto the first binomial order. This expression of $h(R,\mathcal{G})$ immediately transforms Eq.(\ref{FT eom2}) as
\begin{eqnarray}
 \tilde{A}_r''(k,\eta) + \bigg(k^2 - \frac{4B\epsilon q}{\eta_0^{2q}\big(1 - \frac{2B}{\eta_0^{2q}}\big)}\frac{1}{\eta^2}\bigg)\tilde{A}_r(k,\eta) = 0,
 \label{FT eom3}
\end{eqnarray}
and solving it for $\tilde{A}_r(k,\eta)$, we get
\begin{eqnarray}
 \tilde{A}_r(k,\eta) = \sqrt{-k\eta}\bigg[D_1~J_{\nu}(-k\eta) + D_2~J_{-\nu}(-k\eta)\bigg],
 \label{sol1}
\end{eqnarray}
where $\nu^2 = \frac{1}{4} - \frac{4B\epsilon q}{\eta_0^{2q}\big(\frac{2B}{\eta_0^{2q}} - 1\big)}$, $J_{\nu}$ is the Bessel function of the first kind 
and it may be observed that the mode function for both polarizations has the same solution. 
The solution of Eq.(\ref{FT eom3}) can also be expressed in terms of $J_{\nu}$ and $Y_{\nu}$ where $Y_{\nu}$ is the modified Bessel function, however here 
it is advantageous to replace the modified Bessel function by $J_{\nu}$ and $J_{-\nu}$ and thus the mode function solution has the form of 
Eq.(\ref{sol1}). Moreover, $D_1$, $D_2$ are two integration constants 
which can be further determined from the initial condition of $\tilde{A}_r(k,\eta)$ and as an initial state of the mode function, we consider the 
Bunch-Davies vacuum. Actually, in the deep sub-Hubble regime, i.e in the regime of $|k\eta| \gg 1$, the Mukhanov-Sasaki equation can be 
approximated to $\tilde{A}_r''(k,\eta) + k^2\tilde{A}_r(k,\eta) = 0$, which possesses the Bunch-Davies 
solution like $\tilde{A}_r(k,\eta) = \frac{1}{\sqrt{2k}}e^{-ik\eta}$ and, thus, the Bunch-Davies initial condition is well justified 
in the present context. In the sub-Hubble region of the distant past, the Bessel functions have the following limit 
\begin{eqnarray}
 \lim_{|k\eta| \gg 1} J_{\nu}(-k\eta)&=&\sqrt{\frac{2}{\pi(-k\eta)}}~\cos{\bigg[-k\eta - \frac{\pi}{2}\big(\nu + \frac{1}{2}\big)\bigg]}\nonumber\\
 \lim_{|k\eta| \gg 1} J_{-\nu}(-k\eta)&=&\sqrt{\frac{2}{\pi(-k\eta)}}~\sin{\bigg[-k\eta + \frac{\pi}{2}\big(\nu + \frac{1}{2}\big)\bigg]},
 \nonumber
\end{eqnarray}
and plugging the above expressions into Eq.(\ref{sol1}), we get the sub-Hubble limit of the electromagnetic mode function as
\begin{eqnarray}
 \lim_{|k\eta| \gg 1} \tilde{A}_r(k,\eta)&=&\sqrt{\frac{2}{\pi}}\bigg[D_1~\cos{\big[-k\eta - \frac{\pi}{2}\big(\nu + \frac{1}{2}\big)\big]} 
 + D_2~\sin{\big[-k\eta + \frac{\pi}{2}\big(\nu + \frac{1}{2}\big)\big]}\bigg]\nonumber\\
 &=&\sqrt{\frac{2}{\pi}}\bigg[e^{-ik\eta}\bigg(\frac{D_1}{2}e^{-i\frac{\pi}{2}(\nu + \frac{1}{2})} 
 + \frac{D_2}{2i}e^{i\frac{\pi}{2}(\nu + \frac{1}{2})}\bigg) + e^{ik\eta}\bigg(\frac{D_1}{2}e^{i\frac{\pi}{2}(\nu + \frac{1}{2})} 
 - \frac{D_2}{2i}e^{-i\frac{\pi}{2}(\nu + \frac{1}{2})}\bigg)\bigg],
 \label{subhorizon form}
\end{eqnarray}
where in the second line we expand the vector potential in terms of $\exp{\big(\pm ik\eta\big)}$. 
Owing to the Bunch-Davies initial condition, the coefficient of the positive frequency mode function is given by $\frac{1}{\sqrt{2k}}$ and that 
of the negative frequency mode function becomes zero. This leads to the following forms of $D_1$ and $D_2$ 
\begin{eqnarray}
 D_1 = \frac{1}{2}\sqrt{\frac{\pi}{k}}~\frac{e^{-i\frac{\pi}{2}(\nu + \frac{1}{2})}}{\cos{\big[\pi(\nu + 1/2)\big]}}~~~~~~~~~~~~,~~~~~~~~~~~
 D_2 = \frac{1}{2}\sqrt{\frac{\pi}{k}}~\frac{e^{i\frac{\pi}{2}(\nu + \frac{3}{2})}}{\cos{\big[\pi(\nu + 1/2)\big]}},
 \label{integration constants}
\end{eqnarray}
and, consequently, the final solution of the mode function becomes
\begin{eqnarray}
 A_r(k,\eta)&=&\frac{1}{h(\eta)}\tilde{A}_r(k,\eta)\nonumber\\
 &=&\frac{\sqrt{-k\eta}}{2\bigg[1 - \frac{4B}{\eta_0^{2q}}\bigg(\frac{-\eta}{\eta_0}\bigg)^{2\epsilon q}\bigg]^{1/2}}~
 \bigg\{\sqrt{\frac{\pi}{k}}~\frac{e^{-i\frac{\pi}{2}(\nu + \frac{1}{2})}}{\cos{\big[\pi(\nu + 1/2)\big]}}~J_{\nu}(-k\eta) 
 + \sqrt{\frac{\pi}{k}}~\frac{e^{i\frac{\pi}{2}(\nu + \frac{3}{2})}}{\cos{\big[\pi(\nu + 1/2)\big]}}~J_{-\nu}(-k\eta)\bigg\}~~~.
 \label{final solution}
\end{eqnarray}
In the superhorizon limit, in which case the modes are outside of the Hubble radius i.e $k < \frac{1}{\mathcal{H}}$ 
(recall $\mathcal{H}$ is the Hubble parameter), the mode function can be expressed as 
\begin{eqnarray}
 \lim_{|k\eta| \ll 1} A_r(k,\eta) = \frac{1}{\bigg[1 - \frac{4B}{\eta_0^{2q}}\bigg(\frac{-\eta}{\eta_0}\bigg)^{2\epsilon q}\bigg]^{1/2}}~
 \bigg\{\frac{D_1}{2^{\nu}\Gamma(\nu + 1)}~\big(-k\eta\big)^{\nu + \frac{1}{2}} + \frac{D_2}{2^{-\nu}\Gamma(-\nu + 1)}
 ~\big(-k\eta\big)^{-\nu + \frac{1}{2}}\bigg\},
 \label{superhorizon form}
\end{eqnarray}
where we have used the power law expansion of the Bessel function given by 
$\lim_{|k\eta| \ll 1} J_{\nu}(-k\eta) = \frac{1}{2^{\nu}\Gamma(\nu + 1)}(-k\eta)^{\nu}$, and similarly for 
$J_{-\nu}(-k\eta)$. Eq.(\ref{superhorizon form}) clearly indicates that $A_r(k,\eta)$ depends on 
the parameter $q$ and the inflationary energy scale determined by $\eta_0^{-1}$ (recall $\eta_0^{-1} = H_0$ actually 
specifies the cosmic Hubble parameter at the 
beginning of inflation, as in Eq.(\ref{cosmic Hubble parameter})). 
As mentioned earlier, $q$ is considered to lie within $0 < q < 1$ and, moreover, $H_0$ is generally 
considered to be $H_0 = 10^{-5}M_{Pl} \approx 10^{14}\mathrm{GeV}$. Having derived the vector potential, let us now turn to the electric and magnetic power spectrum and their 
respective scale dependencies on $k$. For this purpose, first we need to recall, 
$\nu^2 = \frac{1}{4} - \frac{4B\epsilon q}{\eta_0^{2q}\big(2B/\eta_0^{2q} - 1\big)}$ with 
$B = \kappa^{2q}\big\{\big[6\beta(\beta + 1)\big]^q + \big[-24(\beta + 1)^3\big]^{q/2}\big\}$; which depicts that $\nu$ must be positive as the parameter 
$q$ is greater than zero. Thereby, in the superhorizon limit, it is evident that 
$(-k\eta)^{\nu + \frac{1}{2}} \ll (-k\eta)^{-\nu + \frac{1}{2}}$, and thus the term containing $D_1$ in the right hand side of Eq.(\ref{superhorizon form}) 
is negligible, as compared to the other one containing $D_2$. Consequently, from Eq.(\ref{power spectra}) we have the following power spectra 
for electric and magnetic fields as,
\begin{eqnarray}
 \frac{\partial \rho(\vec{E})}{\partial \ln{k}} = \frac{P(\eta)~(\nu - \frac{1}{2})^2~H_0^4}{4\pi~
 \bigg[1 - \frac{4B}{\eta_0^{2q}}\bigg(\frac{-\eta}{\eta_0}\bigg)^{2\epsilon q}\bigg]~\bigg\{2^{-\nu}\Gamma(-\nu + 1)\cos{[\pi(\nu + 1/2)]}\bigg\}^2} 
 \big(-k\eta\big)^{3-2\nu}
 \label{electric power spectrum 1}
\end{eqnarray}
and
 \begin{eqnarray}
 \frac{\partial \rho(\vec{B})}{\partial \ln{k}} = \frac{Q(\eta)~H_0^4}{4\pi~
 \bigg[1 - \frac{4B}{\eta_0^{2q}}\bigg(\frac{-\eta}{\eta_0}\bigg)^{2\epsilon q}\bigg]~\bigg\{2^{-\nu}\Gamma(-\nu + 1)\cos{[\pi(\nu + 1/2)]}\bigg\}^2} 
 \big(-k\eta\big)^{5-2\nu},
 \label{magnetic power spectrum 1}
\end{eqnarray}
respectively. The above two expressions 
help us to explore the possibility of a scale invariant magnetic field spectrum. We may note that the scale 
invariance for the magnetic field spectrum does not imply the scale invariance of the electric field spectrum, in particular the electric power spectrum becomes 
scale invariant for $\nu = 3/2$ while, in the magnetic case, $\nu = 5/2$ leads to a scale invariant power spectrum. However one can 
easily investigate that for any possible value of $q$, i.e.. within $0 < q < 1$, 
the function $\nu(q)$ does not acquire either the value $\frac{3}{2}$ or $\frac{5}{2}$ and thus the electric and magnetic 
power spectra are $not~scale~invariant$ in the present context. Using the expression of $\nu$ (given just after Eq.(\ref{sol1})), we give the plot 
of $\nu(q)$ versus $q$ in Fig.[\ref{plot1}], which clearly demonstrates that $\nu(q)$ lies below  the values required 
for a scale invariant electric or magnetic field spectrum (i.e.  values like $\frac{3}{2}$ or $\frac{5}{2}$ respectively). 
This leads to the aforementioned argument of the impossibility of having both
a scale invariant electric as well as  
magnetic power spectrum, in the present scenario; in particular, the non-zero electric and magnetic spectral indexes are given by 
$n(E) = 3-2\nu$ and $n(B) = 5-2\nu$, respectively. Moreover Fig.[\ref{plot1}] also reveals that the spectral indexes are positive, which hints to the 
resolution of the backreaction problem.

However, in order to ensure that the backreaction of the electromagnetic field stays small during inflation, 
we shall first consider the energy stored in the electric field at a given time $\eta_c$, which is 
\begin{eqnarray}
 \rho(\vec{E},\eta_c) = \int_{k_i}^{k_c} \frac{\partial \rho(\vec{E})}{\partial \ln{k}} d\ln{k} 
 = \frac{P(\eta_c)~(\nu - \frac{1}{2})^2~H_0^4~\big(1 - e^{-2N_c}\big)}{4\pi~
 \bigg[1 - \frac{4B}{\eta_0^{2q}}\bigg(\frac{-\eta_c}{\eta_0}\bigg)^{2\epsilon q}\bigg]~\bigg\{2^{-\nu}\Gamma(-\nu + 1)\cos{[\pi(\nu + 1/2)]}\bigg\}^2},
 \label{electric energy density 1}
\end{eqnarray}
where we used Eq.(\ref{electric power spectrum 1}), 
$k_i$ and $k_c$ denote the mode-momenta that cross the horizon at the beginning of inflation and at $\eta = \eta_c$, respectively, and thus $k_i$ 
is safely considered to be the same as for the CMB scale. Moreover, $N_c$ 
is the number of inflationary e-foldings up to the time $\eta = \eta_c$ and, thereby, $N_c$ is given by $N_c = \ln{\big(\frac{a_c}{a_i}\big)}$, with 
$a_i = a(\eta_i)$ and $a_c = a(\eta_c)$. We also have the relation $|\eta_c| = k_c^{-1}$ and $k_c$ is obviously greater than the CMB scale 
momentum $k_{CMB} \approx 10^{-40}\mathrm{GeV} \approx 0.02 \mathrm{Mpc}^{-1}$. Similarly, using Eq.(\ref{magnetic power spectrum 1}), we determine 
the energy density coming from the magnetic fields, which yields
\begin{eqnarray}
 \rho(\vec{B},\eta_c) = \int_{k_i}^{k_c} \frac{\partial \rho(\vec{B})}{\partial \ln{k}} d\ln{k} 
 = \frac{Q(\eta_c)~H_0^4~\big(1 - e^{-4N_c}\big)}{4\pi~
 \bigg[1 - \frac{4B}{\eta_0^{2q}}\bigg(\frac{-\eta_c}{\eta_0}\bigg)^{2\epsilon q}\bigg]~\bigg\{2^{-\nu}\Gamma(-\nu + 1)\cos{[\pi(\nu + 1/2)]}\bigg\}^2}~~.
 \label{magnetic energy density 1}
\end{eqnarray}
The total electromagnetic energy density at $\eta = \eta_c$ becomes
\begin{eqnarray}
 \rho_{em}(\eta_c) = \rho(\vec{E},\eta_c) + \rho(\vec{B},\eta_c) = 
 \frac{H_0^4\bigg[P(\eta_c)~(\nu - \frac{1}{2})^2~\big(1 - e^{-2N_c}\big) + Q(\eta_c)~\big(1 - e^{-4N_c}\big)\bigg]}{4\pi~
 \bigg[1 - \frac{4B}{\eta_0^{2q}}\bigg(\frac{-\eta_c}{\eta_0}\bigg)^{2\epsilon q}\bigg]~\bigg\{2^{-\nu}\Gamma(-\nu + 1)\cos{[\pi(\nu + 1/2)]}\bigg\}^2}.
 \label{total energy density}
\end{eqnarray}
where $P(\eta)$ and $Q(\eta)$ are shown in Eqs.(\ref{P}) and (\ref{Q}), respectively. In order to avoid the backreaction issue, we have to ensure that the 
electromagnetic energy density is less than that of the background energy density; in particular, we have to show $\rho_{em} < 3M_{Pl}^2H^2$ during  
inflation. In the context of the scalar-Einstein-Gauss-Bonnet theory, which is considered to be the background gravity theory in the present work, the background energy density gets
 contributions from the scalar field and also from the higher curvature terms through the GB coupling function. In regard to the explicit expressions 
of the functions $P(\eta)$ and $Q(\eta)$, plugging the conformal Hubble parameter from Eq.(\ref{Hubble parameter}) to Eqs.(\ref{P}), (\ref{Q}) and 
after simplifying a bit, we obtain
\begin{eqnarray}
 P(\eta_c)&=&1 - 4\big(1 + \frac{q}{2}\big)\big(12^q + 24^{q/2}\big)\bigg(\frac{H_0}{M_\mathrm{Pl}}\bigg)^{2q}\bigg(-\frac{\eta_c}{\eta_0}\bigg)^{2\epsilon q}\nonumber\\
 Q(\eta_c)&=&1 - 4\big(1 - \frac{q}{2}\big)\big(12^q + 24^{q/2}\big)\bigg(\frac{H_0}{M_\mathrm{Pl}}\bigg)^{2q}\bigg(-\frac{\eta_c}{\eta_0}\bigg)^{2\epsilon q},
 \label{P and Q modified}
\end{eqnarray}
where we used $\eta_0^{-1} = H_0$. 
The above expressions, along with the consideration of $0 < q < 1$ and $\epsilon < 1$, Eq.(\ref{total energy density}), clearly indicate that the 
electromagnetic energy density during inflation is of the order of $H_0^4$, i.e.
\begin{eqnarray}
 \rho_{em}(\eta_c) \sim H_0^4~~~.
 \label{backreaction}
\end{eqnarray}
The inflationary energy scale is less than the Planck scale; in particular, we consider $H_0 = 10^{-5}M_{Pl}$ and, thus, Eq.(\ref{backreaction}) leads to the 
inequality  $\rho_{em} \ll M_{Pl}^2H^2$. This confirms that the electromagnetic field has a negligible backreaction 
on the background inflationary spacetime, leading to the resolution of the backreaction problem in the present magnetogenesis scenario.

\section{Curvature perturbation sourced by EM field during inflation}

The production of the gauge field during inflation may source the curvature perturbation in the super Hubble scales 
\cite{Fujita:2013qxa,Fujita:2016qab,Barnaby:2012tk,Ferreira:2014hma,Giovannini:2013rme,Bamba:2014vda,Suyama:2012wh}. The power spectrum of the induced curvature 
perturbations or the induced non-Gaussianities should satisfy the Planck constraints. Thereby in the present context where the EM field 
non-minimally couples with the background spacetime curvature, it is important to investigate whether the curvature perturbations induced by the EM field 
obeys the Planck constraints or not. Earlier, in \cite{Fujita:2013qxa,Fujita:2016qab,Barnaby:2012tk,Ferreira:2014hma,
Giovannini:2013rme,Bamba:2014vda,Suyama:2012wh}, 
the authors discussed such kind of induced curvature perturbations and the corresponding 
constraints on the model parameters in a magnetogenesis scenario where the EM field couples with a scalar field (say, the inflaton field). However 
in the present paper, the EM field couples with the background Ricci scalar and the Gauss-Bonnet invariant, unlike to 
the case where the EM field couples with a scalar field.

The curvature perturbation $\zeta(\eta,\vec{x})$ is defined as the perturbation of the scale factor $a(\eta,\vec{x})$ on 
the uniform density slice, i.e $\zeta(\eta,\vec{x}) = \ln{[a(\eta,\vec{x})/a(\eta)]}$ where $\eta$ is the cosmic time. Then the curvature perturbation that is 
sourced by the EM field is given by \cite{Fujita:2013qxa},
\begin{eqnarray}
 \zeta_{em}(\eta,\vec{x}) = -\frac{2H}{\epsilon\rho_{inf}}\int_{\eta_m}^{\eta}\frac{d\eta'}{a(\eta')} \rho_{em}(\eta',\vec{x})
 \label{induced1}
\end{eqnarray}
where $H$ is the cosmic time Hubble parameter during inflation, $\epsilon$ is the slow roll parameter, 
$\rho_{inf}$ is the background inflaton energy density 
(recall, the scalar coupled Gauss-Bonnet curvature is responsible for the inflation in the present context) and $\rho_{em}$ denotes the EM field 
energy density. The lower limit $\eta_m$ in the integral corresponds to the time after which the gauge field production starts and the mode which 
crosses the horizon at $\eta = \eta_m$ will be symbolized by $k_{min}$ in the later calculation. Here we assume $k_{CMB} > k_{min}$, i.e the 
generation of electromagnetic fields is considered to begin earlier than the horizon-crossing of CMB modes. At this stage it deserves mentioning that 
the electromagnetic anisotropic stress which can also source the curvature perturbation is not taken into account in Eq.(\ref{induced1}). 
This is due to the fact that the contribution from the electromagnetic anisotropic stress is suppressed during slow roll inflation, in particular 
by the inverse of the slow roll parameter $\epsilon$, in comparison to the contribution written in the right hand side of Eq.(\ref{induced1}) 
\cite{Suyama:2012wh}. 
The EM field energy density can 
be expressed as $\rho_{em}(\eta,\vec{x}) = \rho_E(\eta,\vec{x}) + \rho_B(\eta,\vec{x})$, however Eqs.(\ref{electric power spectrum 1}) and 
(\ref{magnetic power spectrum 1}) clearly indicate that the ratio of magnetic to electric power spectrum in the superhorizon scale is given by: 
$\mathcal{P}_B/\mathcal{P}_E \sim \big(-k\eta\big)^2/\big(\nu - \frac{1}{2}\big)^2$ which depicts that the magnetic field strength is much lower 
than the electric strength in the superhorizon limit. Thereby we can consider the EM field energy density as 
$\rho_{em} \approx \rho_E = \frac{1}{2}E^2$. Such consideration allows to express the EM field energy density in Fourier space as follows,
\begin{eqnarray}
 \rho_{em}(\eta,\vec{k}) = \frac{1}{2}\int\int \frac{d^3p_1d^3p_2}{(2\pi)^3}~\delta\big(\vec{p}_1 + \vec{p}_2 - \vec{k}\big) 
 \vec{E}(\eta,\vec{p}_1)\vec{E}(\eta,\vec{p}_2)~~~,
 \label{induced2}
\end{eqnarray}
where the electric field is defined as $\big|E(\eta,k)\big| = \frac{1}{a^2}\big|A'(k,\eta)\big|$. Using Eq.(\ref{superhorizon form}), we determine the 
electric field during inflation as,
\begin{eqnarray}
 \big|E(\eta,k)\big| = \frac{\big(\nu - \frac{1}{2}\big)}{\sqrt{2}}H^2k^{-\nu}\big(-\eta\big)^{\frac{3}{2} - \nu}
 \label{induced3}
\end{eqnarray}
with, recall, $\nu^2 = \frac{1}{4} - \frac{4B\epsilon q}{\eta_0^{2q}\big(2B/\eta_0^{2q} - 1\big)}$ and 
$B = \kappa^{2q}\big\{\big[6\beta(\beta + 1)\big]^q + \big[-24(\beta + 1)^3\big]^{q/2}\big\}$. Using the above expression of electric field and 
following the procedure of \cite{Fujita:2013qxa}, we evaluate the 2-point correlator of $\zeta_{em}(\eta,\vec{k})$ at $\eta = \eta_f$ 
(i.e at the end of inflation) in the present context as,
\begin{eqnarray}
 \langle\zeta_{em}(\eta_f,\vec{k}_1)\zeta_{em}(\eta_f,\vec{k}_2)\rangle&=&2\delta\big(\vec{k}_1 + \vec{k}_2\big) G_2 
 \int_{k_{min}}^{k_{f}} d^3p_1d^3p_2 \delta\big(\vec{p}_2 - \vec{p}_1 - \vec{k}_2\big) p_1^{-2\nu}p_2^{-2\nu}\nonumber\\
 &\bigg(&\delta_{j_1j_2} - \big(\hat{p}_1\big)_{j_1}\big(\hat{p}_1\big)_{j_2}\bigg)\bigg(\delta_{j_1j_2} - \big(\hat{p}_2\big)_{j_1}\big(\hat{p}_2\big)_{j_2}\bigg) 
 \times\bigg\{\prod_{i=1,2}~\int_{\eta_m}^{\eta_f}d\eta_i\big(-\eta_i\big)^{2-2\nu}\bigg\}
 \label{induced4}
\end{eqnarray}
where $k_{f}$ is the mode that crosses the horizon at the end of inflation i.e at $\eta = \eta_f$ and $k_{min}$ is defined after Eq.(\ref{induced1}). 
Moreover the factor $G_2$, present in the above expression, is given by,
\begin{eqnarray}
 G_2 = \bigg[\frac{H_f^2\big(\nu - \frac{1}{2}\big)^2}{6\epsilon M_{Pl}^2}\bigg]^2~~.
 \label{G}
\end{eqnarray}
To derive $G_2$, we use $\rho_{inf}(\eta_f) = 3H_f^2M_{Pl}^2$ which holds true due to the fact that the EM field has negligible backreaction 
on the background inflationary spacetime. Performing the $p_2$ and the $\eta$ integral of Eq.(\ref{induced4}), we get
\begin{eqnarray}
 \langle\zeta_{em}(\eta_f,\vec{k}_1)\zeta_{em}(\eta_f,\vec{k}_2)\rangle = \frac{32\pi}{3}\delta\big(\vec{k}_1 + \vec{k}_2\big) G_2
 \int_{k_{min}}^{k_f} dp_1~p_1^{2-2\nu}\big(p_1 + k_2\big)^{-2\nu} \bigg\{\frac{\big(-\eta_f\big)^{3-2\nu} - \big(-\eta_m\big)^{3-2\nu}}{3-2\nu}\bigg\}^2~.
 \label{induced5}
\end{eqnarray}
where we use the integral $\int d\Omega_k\hat{k}_i\hat{k}_j = \frac{4\pi}{3}\delta_{ij}$. 
Here we would like to mention that the quantity $\nu = \frac{1}{2}\bigg[1 - \frac{16B\epsilon q}{\eta_0^{2q}\big(2B/\eta_0^{2q} - 1\big)}\bigg]^{1/2}$ is 
less than $1/2$, which can also be ensured from the Fig.[\ref{plot1}]. Thereby the integral in Eq.(\ref{induced5}) will get the maximum 
contribution from the upper limit $k_{f}$ and as a result, the final form of the two point correlator comes as,
\begin{eqnarray}
 \langle\zeta_{em}(\vec{k}_1)\zeta_{em}(\vec{k}_2)(\eta_f)\rangle = \frac{32\pi}{3k_1^3}\delta\big(\vec{k}_1 + \vec{k}_2\big) G_2
 \bigg\{\frac{\big(k_{f}/k_1\big)^{3-4\nu}}{3-2\nu}\bigg\} \bigg\{\frac{\big(-k_1\eta_f\big)^{3-2\nu} - \big(-k_1\eta_m\big)^{3-2\nu}}{3-2\nu}\bigg\}^2~.
 \label{2-point}
\end{eqnarray}
We will eventually consider the two point correlator at the CMB scales, i.e $k_1 = k_{CMB}$, and since, as mentioned earlier, the EM field generation 
starts earlier than the horizon-crossing of the CMB modes, we have $k_{f} \gg k_1 = k_{CMB} \gg k_{min}$. Furthermore, by using Eq.(\ref{induced3}) 
along with the described procedure in \cite{Fujita:2013qxa}, we calculate the 3-point and 4-point correlators of the induced curvature perturbation and they 
are given by the following expressions,
\begin{eqnarray}
 \langle\zeta_{em}(\vec{k}_1)\zeta_{em}(\vec{k}_2)\zeta_{em}(\vec{k}_3)(\eta_f)\rangle&=&
 \frac{32\pi}{3}\delta\big(\vec{k}_1 + \vec{k}_2 + \vec{k}_3\big) G_3
 \bigg\{\frac{\big(k_{f}/k_1\big)^{3-6\nu}}{3-2\nu}\bigg\} \bigg\{\frac{\big(-k_1\eta_f\big)^{3-2\nu} - \big(-k_1\eta_m\big)^{3-2\nu}}{3-2\nu}\bigg\}^3\nonumber\\ 
 &\bigg\{&\frac{1 + \big(\hat{k}_1.\hat{k}_2\big)^2}{\big(k_1k_2\big)^3} + \frac{1 + \big(\hat{k}_1.\hat{k}_3\big)^2}{\big(k_1k_3\big)^3} 
 + \frac{1 + \big(\hat{k}_2.\hat{k}_3\big)^2}{\big(k_2k_3\big)^3}\bigg\}~.
 \label{3-point}
\end{eqnarray}
and
\begin{eqnarray}
 \langle\zeta_{em}(\vec{k}_1)\zeta_{em}(\vec{k}_2)\zeta_{em}(\vec{k}_3)\zeta_{em}(\vec{k}_4)(\eta_f)\rangle&=&
 \frac{64\pi}{3}\delta\big(\vec{k}_1 + \vec{k}_2 + \vec{k}_3 + \vec{k}_4\big) G_4
 \bigg\{\frac{\big(k_{f}/k_1\big)^{3-8\nu}}{3-2\nu}\bigg\} \bigg\{\frac{\big(-k_1\eta_f\big)^{3-2\nu} - \big(-k_1\eta_m\big)^{3-2\nu}}{3-2\nu}\bigg\}^4\nonumber\\ 
 &\bigg\{&\frac{\big(\hat{k}_1.\hat{k}_2\big)^2 + \big(\hat{k}_1.\hat{k}_{13}\big)^2 + \big(\hat{k}_2.\hat{k}_{13}\big)^2 - 
 \big(\hat{k}_1.\hat{k}_2\big)\big(\hat{k}_1.\hat{k}_{13}\big)\big(\hat{k}_2.\hat{k}_{13}\big)}{\big(k_1k_2k_{13}\big)^3} + 11~\mathrm{perms.}\bigg\}
 \label{4-point}
\end{eqnarray}
respectively, with $\vec{k}_{13} = \vec{k}_1 + \vec{k}_3$. Moreover $G_3$ and $G_4$ have the following forms,
\begin{eqnarray}
 G_3 = \bigg[\frac{H_f^2\big(\nu - \frac{1}{2}\big)^2}{6\epsilon M_{Pl}^2}\bigg]^3~~~~~~~,~~~~~~
 G_4 = \bigg[\frac{H_f^2\big(\nu - \frac{1}{2}\big)^2}{6\epsilon M_{Pl}^2}\bigg]^4~~.
 \label{G3 and G4}
\end{eqnarray}
The 2-point correlator in Eq.(\ref{2-point}) immediately leads to the power spectrum of the curvature perturbation induced by the EM field as,
\begin{eqnarray}
 \mathcal{P}(\zeta_{em}) = \frac{2}{3}\bigg[\frac{H_f^2\big(\nu - \frac{1}{2}\big)^2}{6\epsilon M_{Pl}^2}\bigg]^2
 \bigg\{\frac{\big(k_{f}/k_1\big)^{3-4\nu}}{3-2\nu}\bigg\} \bigg\{\frac{\big(-k_1\eta_f\big)^{3-2\nu} - \big(-k_1\eta_m\big)^{3-2\nu}}{3-2\nu}\bigg\}^2~~.
 \label{induced power spectrum}
\end{eqnarray}
Similarly the 3-point and 4-point correlators provide 
the induced non-linear parameters $f^{em}_{NL}$ and $\tau_{NL}^{em}$ in the present magnetogenesis model as,
\begin{eqnarray}
 f^{em}_{NL} = \frac{10}{27}\bigg[\frac{H_f^2\big(\nu - \frac{1}{2}\big)^2}{6\epsilon M_{Pl}^2}\bigg]^3\bigg(\frac{1}{\mathcal{P}(\zeta_{em})}\bigg)^2
 \bigg\{\frac{\big(k_{f}/k_1\big)^{3-6\nu}}{3-2\nu}\bigg\} \bigg\{\frac{\big(-k_1\eta_f\big)^{3-2\nu} - \big(-k_1\eta_m\big)^{3-2\nu}}{3-2\nu}\bigg\}^3~~.
 \label{NL1}
\end{eqnarray}
and
\begin{eqnarray}
 \tau^{em}_{NL} = \frac{1}{3}\bigg[\frac{H_f^2\big(\nu - \frac{1}{2}\big)^2}{6\epsilon M_{Pl}^2}\bigg]^4\bigg(\frac{1}{\mathcal{P}(\zeta_{em})}\bigg)^3
 \bigg\{\frac{\big(k_{f}/k_1\big)^{3-8\nu}}{3-2\nu}\bigg\} \bigg\{\frac{\big(-k_1\eta_f\big)^{3-2\nu} - \big(-k_1\eta_m\big)^{3-2\nu}}{3-2\nu}\bigg\}^4~~.
 \label{NL2}
\end{eqnarray}
respectively. As mentioned earlier, the mode $k_1$ is identified with the CMB scale and thus we have the relations like 
$-k_1\eta_f = e^{-N_f}$ (where $N_f$ is the e-fold number from the horizon crossing of $k_{CMB}$ to the horizon-crossing of $k_f$, i.e $N_f$ is the 
inflationary e-fold number) and $-k_1\eta_m = e^{\big(N_{em} - N_f\big)}$ where $N_{em}$ is the e-fold from the horizon crossing of $k_{min}$ to the 
horizon crossing of $k_f$, i.e $N_{em}$ denotes the e-fold during which the EM field production occurs. Eqs.(\ref{induced power spectrum}), 
(\ref{NL1}) and (\ref{NL2}) clearly reveal that $\mathcal{P}(\zeta_{em})$, $f^{em}_{NL}$ and $\tau^{em}_{NL}$ depend on the quantities: 
$H_0/M_{Pl}$, $H_f/M_{Pl}$, $\epsilon$, $q$, $N_f$ and $N_{em} - N_f$. Out of these quantities, $H_f$ is related to $H_0$, $\epsilon$ and 
$N_f$ via $H_f = H_0\exp{\big[\frac{-\epsilon N_f}{1+\epsilon}\big]}$, as shown in Eq.(\ref{cosmic Hubble parameter}). 
In particular, $H_0 = 10^{-5}M_{Pl}$, $\epsilon = 0.1$, $N_f = 58$ immediately leads to the Hubble parameter at the end of inflation as 
$H_f = 5.1\times10^{-8}M_{Pl}$. Due to such parametric regime along with $q = 0.5$ (which we will also consider in the next section during the determination of 
magnetic field's strength at present epoch), Eqs.(\ref{induced power spectrum}), (\ref{NL1}) and (\ref{NL2}) become,
\begin{eqnarray}
 \mathcal{P}(\zeta_{em})&\approx&10^{-27}\times\exp{\big[4\big(N_{em} - N_f\big)\big]}\nonumber\\
 f^{em}_{NL}&\approx&10^{-24}\times\exp{\big[6\big(N_{em} - N_f\big)\big]}\nonumber\\
 \tau^{em}_{NL}&\approx&10^{-23}\times\exp{\big[8\big(N_{em} - N_f\big)\big]}
 \label{estimation}
\end{eqnarray}
Having the theoretical predictions of $\mathcal{P}(\zeta_{em})$, $f^{em}_{NL}$ and $\tau^{em}_{NL}$ in hand, we now confront the model with the Planck 
results which put certain constraints on such quantities given by,
\begin{eqnarray}
 \mathcal{P}^{obs}(\zeta) \approx 2.1\times10^{-9}~~~,~~~~
 f_{NL} \leq 14.3 = f_{NL}^{obs} (\mathrm{say})~~~,~~~~
 \tau_{NL} \leq 2800 = \tau_{NL}^{obs} (\mathrm{say})~~.
 \label{constraints}
\end{eqnarray}
The restriction, that the theoretical predictions of 
$\mathcal{P}(\zeta_{em})$, $f^{em}_{NL}$ and $\tau^{em}_{NL}$ do not exceed their respective observed values provided by the Planck 
results, in turn lead to corresponding constraint on $N_{em} - N_f$, in particular,
\begin{eqnarray}
 \mathcal{P}(\zeta_{em}) < \mathcal{P}^{obs}(\zeta) &\Longrightarrow& N_{em} - N_f < \frac{1}{4}\ln{\bigg[10^{27}\times\mathcal{P}^{obs}(\zeta)\bigg]} 
 \approx 10.4\nonumber\\
 f^{em}_{NL} < f_{NL}^{obs} &\Longrightarrow& N_{em} - N_f < \frac{1}{6}\ln{\bigg[10^{24}\times f_{NL}^{obs}\bigg]} \approx 9.6\nonumber\\
 \tau^{em}_{NL} < \tau_{NL}^{obs} &\Longrightarrow& N_{em} - N_f < \frac{1}{8}\ln{\bigg[10^{23}\times \tau_{NL}^{obs}\bigg]} \approx 7.6
 \label{constraint Nem}
\end{eqnarray}
Eq.(\ref{constraint Nem}) clearly evidents that the allowed space of $N_{em} - N_f$ becomes tighter due to the restriction 
$\tau^{em}_{NL} < \tau_{NL}^{obs}$ compared to the other two restrictions. 
Thus as a whole, Eq.(\ref{constraint Nem}) provides the constraints from the curvature perturbation induced by 
the electromagnetic field during inflation in the present magnetogenesis scenario where the EM field couples with the background 
Ricci scalar and the Gauss-Bonnet curvature.  

\section{Present magnetic strength for the instantaneous reheating case}\label{sec_instantaneous reheating}
In this section we will concentrate on the strength of the magnetic field in the present epoch, as it is important 
to know whether the model can generate magnetic fields of sufficient strength. For that purpose we need 
to know the conductivity of the universe, both during the inflationary epoch and immediately after it. During inflation the 
universe was a poor electrical conductor, however after inflation it became 
a very good conductor and hence the electric currents became important during this phase. 
The high electrical conductivity in the post inflationary epoch lies on 
the consideration that at the end of inflation, we assume $instantaneous~reheating$ (the reheating e-folding number is taken to be zero), 
i.e., the universe makes a sudden 
jump from the inflationary phase to a radiation dominated epoch during which the cosmic Hubble parameter goes as $H \propto t^{-1}$ with 
$t$ being the cosmic time (at a later stage, we will relax this  assumption and will consider a reheating epoch with a non-zero e-folding number 
after inflation). Due to the large conductivity in the post inflationary phase, one can write the current density as $J^{i} = \sigma E^{i}$ where 
$\sigma$ is the conductivity and $E^{i}$ the components of the electric field. Then, the corresponding vector potential has two sorts of 
solutions: one independent of time and the other behaving as $\exp{(-\sigma t)}$, which is vanishingly small. 
Thus, the vector potential remains constant with time and suggests that the electric field becomes soon negligible, 
while the magnetic field remains as the dominant piece. Moreover, we mentioned earlier that at late time the spacetime curvature becomes low enough that the 
conformal breaking term $S_{CB}$ will not contribute and the electromagnetic field follows the standard Maxwell's equations. In particular, 
we consider $f(R,\mathcal{G})$ to be zero during the post inflationary phase, which can be also connected with the other
point of view, as Eq.(\ref{form of f 2}) clearly shows that the conformal coupling $f(R,\mathcal{G})$ 
goes to zero (and, consequently, also $P(\eta) = Q(\eta) = 1$) as $k\eta \rightarrow 0$, i.e. at the end of inflation. The conformal symmetry 
of the electromagnetic field is thus restored after inflation, and  the electromagnetic energy density decays as $1/a^4$ or, equivalently, the magnetic field energy density evolves as $1/a^4$, as the electric field is practically zero in the post inflationary epoch. Hence, the magnetic field strength at the 
present epoch is related with that at the end of inflation by the expression
\begin{eqnarray}
 \frac{\partial \rho(\vec{B})}{\partial \ln{k}}\bigg|_{0} = \bigg(\frac{a_f}{a_0}\bigg)^4~\frac{\partial \rho(\vec{B})}{\partial \ln{k}}\bigg|_{{\eta_f}},
 \label{magnetic strength 1}
\end{eqnarray}
where $\eta_f$ is the conformal time at the end of inflation and the suffix '0' denotes present time. With Eq.(\ref{magnetic power spectrum 1}), 
the above equation yields the present magnetic strength ($B_0$) 
\begin{eqnarray}
 B_0 = \frac{1}{\sqrt{2\pi}}~\bigg\{\frac{2^{\nu}}{\Gamma(-\nu + 1)\cos{[\pi(\nu + 1/2)]}~
 \bigg(1 - \frac{4B}{\eta_0^{2q}}\bigg(\frac{-\eta_f}{\eta_0}\bigg)^{2\epsilon q}\bigg)^{1/2}}\bigg\}\bigg(\frac{a_f}{a_0}\bigg)^2~H_0^2\big(-k\eta_f\big)^{-\nu + 5/2},
 \label{magnetic strength 2}
\end{eqnarray}
where we recall that $B = \kappa^{2q}\big\{\big[6\beta(\beta + 1)\big]^q + \big[-24(\beta + 1)^3\big]^{q/2}\big\} = \kappa^{2q}\big(12^q + 24^{q/2}\big)$ 
(appearing in the right hand side of Eq.(\ref{magnetic strength 2})) and $k$ denotes the CMB scale mode-momentum on which we will 
estimate the current magnetic strength. In order to estimate 
$B_0$ from Eq.(\ref{magnetic strength 2}), we need to know $\frac{a_0}{a_f}$ and $k\eta_f$. For the purpose of $\frac{a_0}{a_f}$, we use 
entropy conservation, i.e. $gT^3a^3 = \mathrm{constant}$, where $g$ refers to the effective relativistic degrees of freedom and $T$ is the 
temperature of the relativistic fluid, which finally yields $\frac{a_0}{a_f} \approx 10^{30}\big(H_f/10^{-5}M_{Pl}\big)^{1/2}$, with $H_f$ being the 
Hubble parameter at the end of inflation and can be determined from the corresponding evolution $H = H_0\exp{\big(\frac{-\epsilon N}{1+\epsilon}\big)}$ as 
established in Eq.(\ref{cosmic Hubble parameter}). In particular, considering  $H_0 = 10^{-5}M_{Pl}$, $\epsilon = 0.1$ and the total inflationary 
e-fold number $N_f = 58$, immediately leads to $H_f = 5.1\times10^{-8}M_{Pl}$. Moreover, in regard to 
$k\eta_f$, we have the relation $|\eta_f| = 1/k_{end}$ where $k_{end}$ is the mode which crosses the horizon at the end of inflation and thus 
$k\eta_f$ is given by $-k\eta_f = \exp{(-N_f)}$. With such expressions of $\frac{a_0}{a_f}$ and $k\eta_f$, along with $q = 0.5$, 
we estimate the magnetic strength at the present epoch from Eq.(\ref{magnetic strength 2}) to be
\begin{eqnarray}
 B_0\bigg|_{CMB} \approx 10^{-62}\mathrm{G},
 \label{magnetic strength 3}
\end{eqnarray}
where we use the conversion $1\mathrm{G} = 1.95\times10^{-20}\mathrm{GeV}^2$. 
The above result gives us a typical value for the magnetic field at the present epoch, as obtained from our framework. However, from 
the observational results a constraint on the current magnetic strength of 
$10^{-10}G \lesssim B_0 \lesssim 10^{-22}G$ is obtained around the CMB scales. Therefore the theoretical 
prediction of $B_0$ coming from the present model lies far below  the range  of the observational constraints; and this argument is not just confined 
to $q = 0.5$, but also valid for the whole parametric regime that we consider in the present context, i.e. for $0 < q < 1$.

\section{Present magnetic strength for a Kamionkowski like reheating model with non-zero e-folding number}\label{sec_kamionkowski}
In the case of instantaneous reheating, which we have considered in Sec.[\ref{sec_instantaneous 
reheating}], the conductivity turns on already after the end of 
inflation and, as a result, the electric field quickly goes to zero. However, if we consider the reheating phase with a non-zero e-folding number, then there is no reason 
to keep the assumption that the conductivity becomes large immediately after  inflation. Indeed, the conductivity just remains non-zero and, 
consequently, the strong electric field induces the magnetic field evolution during the epoch between the end of inflation and 
the end of reheating \cite{Kobayashi:2019uqs}. 
This yields less redshift of the magnetic field in the reheating epoch, as compared to $1/a^4$,  and thus the 
magnetic field's present strength may become to be much larger than what has been estimated in Eq.(\ref{magnetic strength 3}). 
In such situation, our next aim 
is to calculate the current magnetic strength in the present magnetogenesis model by considering the reheating phase with a certain, non-zero e-fold number.

Concerning the reheating dynamics, we follow the conventional reheating mechanism given by Kamionkowski et al. \cite{Dai:2014jja}, 
where the inflaton energy is supposed to instantaneously convert into radiation at the end of reheating. 
In this process, the main idea is to parametrize the reheating phase by an effective equation of state $\omega_\mathrm{eff}$, in particular the Hubble parameter 
during reheating is connected to that at the end of inflation by the EoS parameter $\omega_\mathrm{eff}$. Moreover, 
the duration of the reheating phase, characterized by the respective e-fold number $N_\mathrm{re}$, and the reheating temperature ($T_\mathrm{re}$) 
can be expressed in terms of $\omega_\mathrm{eff}$ and of some inflationary parameters by the following relations \cite{Dai:2014jja,Cook:2015vqa},
\begin{eqnarray}
 N_\mathrm{re} = \frac{4}{\big(1 - 3\omega_\mathrm{eff}\big)}
 \bigg[-\frac{1}{4}\ln{\bigg(\frac{45}{\pi^2g_{re}}\bigg)} - \frac{1}{3}\ln{\bigg(\frac{11g_{s,re}}{43}\bigg)} 
 - \ln{\bigg(\frac{k}{a_0T_0}\bigg)} - \ln{\bigg(\frac{(3H_f^2M_\mathrm{Pl}^2)^{1/4}}{H_0}\bigg)} - N_{f}\bigg]~~~,
 \label{reheating e-folding}
\end{eqnarray}
\begin{eqnarray}
 T_\mathrm{re} = H_0\bigg(\frac{43}{11g_{s,re}}\bigg)^{\frac{1}{3}}\bigg(\frac{a_0T_0}{k}\bigg)\exp{\big[-\big(N_f + N_\mathrm{re}\big)\big]}~~,
 \label{reheating temperature}
\end{eqnarray}
where the present CMB temperature is $T_0 = 2.725\mathrm{K}$, 
the pivot scale $\frac{k}{a_0} \approx 0.02\mathrm{Mpc}^{-1}$ (taken as CMB scale) and $a_0$ is the 
present cosmological scale factor. Here, for simplicity, we have taken both the values of the degrees 
of freedom for entropy at reheating (symbolized by $g_{s,re}$) and the effective number of relativistic species upon 
thermalization (symbolized by $g_{re}$) to be the same and of the order of 100, i.e, $g_{s,re} = g_{re} \approx 100$. With this reheating model, 
we are now going to evaluate the electromagnetic mode function and, consequently, the power spectrum during the reheating epoch, in the next subsection.

\subsection{Electromagnetic mode function and power spectrum during the reheating epoch}
After the end of inflation the curvature coupling function $f(R,\mathcal{G})$ is considered to be zero, 
which can be also connected with the continuity 
point of view as Eq.(\ref{form of f 2}) clearly shows that the conformal breaking coupling $f(R,\mathcal{G})$ 
goes to zero as $k\eta \rightarrow 0$, i.e. at the end of inflation. This indicates that the conformal coupling of the 
electromagnetic field is restored in the post inflationary phase and, hence, the evolution of the electromagnetic field becomes 
standard Maxwellian \cite{Kobayashi:2019uqs}. Thus, the gauge field production from the quantum vacuum ceases to exist, in particular the absolute value 
of the Bogoliubov coefficient after inflation becomes constant (with respect to time) and equal to that at the end inflation. 
During the stage between the end of inflation and reheating, the electromagnetic mode function 
will follow the Maxwell equations in vacuum, i.e., the equation of motion (\ref{FT eom1}) with $f(R,\mathcal{G}) = 0$ (or equivalently 
$h(R,\mathcal{G}) = 1$), given by
\begin{eqnarray}
 A''^{(re)}(k,\eta) + k^2A^{(re)}(k,\eta) = 0,
 \label{reheating eom}
\end{eqnarray}
where $A^{(re)}(k,\eta)$ denotes the electromagnetic mode function during reheating and, moreover, recall that 
$A(k,\eta)$ (i.e. without any superscript) symbolizes the electromagnetic mode function in the inflationary phase (see Sec.[\ref{sec_solution_inflation}]). 
Eq.(\ref{reheating eom}), which is free from source term, indicates that we consider the Universe to be a bad conductor during the reheating phase. 
However, due to the Schwinger production, the assumption of zero conductivity demands a proper justification, which we will perform 
in Sec.[\ref{sec_schwinger}]. Solving Eq.(\ref{reheating eom}), one gets
\begin{eqnarray}
 A^{(re)}(k,\eta) = \frac{1}{\sqrt{2k}}\bigg[c_k~e^{-ik(\eta - \eta_f)} + d_k~e^{ik(\eta - \eta_f)}\bigg],
 \label{reheating solution1}
\end{eqnarray}
with $c_k$, $d_k$  two integration constants and $\eta_f$ the end instant of inflation. The integration constants 
can be determined by matching  $A^{(re)}(k,\eta)$ and $A'^{(re)}(k,\eta)$  at the end of inflation; in particular,
\begin{eqnarray}
 A^{(re)}(k,\eta_f = A(k,\eta_f)~~~~~~~~~~~\mathrm{and}~~~~~~~~~~~A'^{(re)}(k,\eta_f) = A'(k,\eta_f),
 \label{reheating continuity1}
\end{eqnarray}
respectively, where $A(k,\eta)$ represents the mode function during inflation and follows Eq.(\ref{final solution}). Therefore, the integration 
constants turn out to be,
\begin{eqnarray}
 c_k&=&\sqrt{\frac{k}{2}}~A(k,\eta_f) + \frac{i}{\sqrt{2k}}~A'(k,\eta_f)\nonumber\\
 d_k&=&\sqrt{\frac{k}{2}}~A(k,\eta_f) - \frac{i}{\sqrt{2k}}~A'(k,\eta_f)
 \label{continuity2}
\end{eqnarray}
with $A(k,\eta_f)$ can be obtained from Eq.(\ref{final solution}) by putting $\eta = \eta_f$, i.e,
\begin{eqnarray}
 A(k,\eta_f) = \frac{\sqrt{-k\eta_f}}{\bigg[1 - \frac{4B}{\eta_0^{2q}}\bigg(\frac{-\eta_f}{\eta_0}\bigg)^{2\epsilon q}\bigg]^{1/2}}~
 \bigg\{D_1~J_{\nu}(-k\eta_f) + D_2~J_{-\nu}(-k\eta_f)\bigg\},
 \label{rehearting continuity3}
\end{eqnarray}
and, consequently, $A'(k,\eta_f)$ is given by
\begin{eqnarray}
 A'(k,\eta_f) = \frac{k}{\bigg[1 - \frac{4B}{\eta_0^{2q}}\bigg(\frac{-\eta_f}{\eta_0}\bigg)^{2\epsilon q}\bigg]^{1/2}}
 \bigg\{\sqrt{-k\eta_f}&\big[&D_1~J_{-1+\nu}(-k\eta_f) + D_2~J_{-1-\nu}(-k\eta_f)\big]\nonumber\\ 
 + \frac{1}{\sqrt{-k\eta_f}}&\big[&D_1\big(\nu - \frac{1}{2}\big)~J_{\nu}(-k\eta_f) - D_2\big(\nu + \frac{1}{2}\big)~J_{-\nu}(-k\eta_f)\big]\bigg\}~~~.
 \label{reheating continuity4}
\end{eqnarray}
Moreover, due to the interaction between the electromagnetic field and the background time dependent FRW spacetime, 
the electromagnetic field vacuum, starting from the Bunch-Davies vacuum at $\eta \rightarrow -\infty$, changes with time and, as a result, particles 
are produced from this  vacuum. Correspondingly, the Bogoliubov coefficients ($\alpha_k(\eta)$ and $\beta_k(\eta)$) 
at time $\eta$ during  reheating are given by 
\begin{eqnarray}
 \alpha_k(\eta)&=&\sqrt{\frac{k}{2}}~A^{(re)}(k,\eta) + \frac{i}{\sqrt{2k}}~A'^{(re)}(k,\eta),\nonumber\\
 \beta_k(\eta)&=&\sqrt{\frac{k}{2}}~A^{(re)}(k,\eta) - \frac{i}{\sqrt{2k}}~A'^{(re)}(k,\eta)~~~~.
 \label{reheating Bogoliubov1}
\end{eqnarray}
With the solution of $A^{(re)}(k,\eta)$, the above expressions boil down to the following
\begin{eqnarray}
 \alpha_k(\eta) = c_k~e^{-ik(\eta - \eta_f)}~~~~~~~~~~~~,~~~~~~~~~~~~\beta_k(\eta) = d_k~e^{ik(\eta - \eta_f)},
 \label{reheating Bogoliubov2}
\end{eqnarray}
which in turn relate $c_k$ and $d_k$ with the Bogoliubov coefficients defined at $\eta = \eta_f$, in particular $c_k = \alpha_k(\eta_f)$ and 
$d_k = \beta_k(\eta_f)$. Therefore, Eq.(\ref{reheating Bogoliubov2}) demonstrates that the absolute value of the Bogoliubov coefficients 
during reheating are time independent and equal to those at the end of inflation. Now, $\big|\beta_k(\eta)\big|$ represents the total number of produced 
particles (having momentum $k$) at time $\eta$ from the Bunch-Davies vacuum defined at $\eta \rightarrow -\infty$. Hence, the time independency of the 
Bogoliubov coefficients during the reheating phase is a direct consequence of the fact that the conformal symmetry of the electromagnetic field is restored 
after inflation. With $c_k = \alpha_k(\eta_f)$ and $d_k = \beta_k(\eta_f)$, Eq.(\ref{reheating solution1}) can be 
alternatively expressed as
\begin{eqnarray}
 A^{(re)}(k,\eta) = \frac{1}{\sqrt{2k}}\bigg[\alpha_k(\eta_f)~e^{-ik(\eta - \eta_f)} + \beta_k(\eta_f)~e^{ik(\eta - \eta_f)}\bigg]~~~.
 \label{reheating solution2}
\end{eqnarray}
By plugging back this solution of $A^{(re)}(k,\eta)$ into Eq.(\ref{power spectra}) and by putting $P(\eta) = Q(\eta) = 1$ (as $f(R,\mathcal{G}) = 0$ in 
the post inflationary phase), we determine the magnetic and electric power spectra during the reheating epoch, as follows
\begin{eqnarray}
 \frac{\partial \rho(\vec{B})}{\partial \ln{k}}&=&\frac{1}{2\pi^2}~\sum_{r=1,2}~\frac{k^5}{a^4}\big|A_r^{(re)}(k,\eta)\big|^2\nonumber\\
 &=&\frac{1}{2\pi^2}\bigg(\frac{k^4}{a^4}\bigg)\bigg[\big|\alpha_k(\eta_f)\big|^2 + \big|\beta_k(\eta_f)\big|^2 
 + 2\big|\alpha_k(\eta_f)~\beta_k(\eta_f)\big|~\cos{\big\{\theta_1 - \theta_2 - 2k(\eta - \eta_f)\big\}}\bigg]
 \label{reheating magnetic power spectrum1}
\end{eqnarray}
and
\begin{eqnarray}
 \frac{\partial \rho(\vec{E})}{\partial \ln{k}}&=&\frac{1}{2\pi^2}~\sum_{r=1,2}~\frac{k^2}{a^4}\big|A_r'^{(re)}(k,\eta)\big|^2\nonumber\\
 &=&\frac{1}{2\pi^2}\bigg(\frac{k^4}{a^4}\bigg)\bigg[\big|\alpha_k(\eta_f)\big|^2 + \big|\beta_k(\eta_f)\big|^2 
 - 2\big|\alpha_k(\eta_f)~\beta_k(\eta_f)\big|~\cos{\big\{\theta_1 - \theta_2 - 2k(\eta - \eta_f)\big\}}\bigg],
 \label{reheating electric power spectrum1}
\end{eqnarray}
respectively, with $\theta_1 = \mathrm{Arg}[\alpha_k(\eta_f)]$ and $\theta_2 = \mathrm{Arg}[\beta_k(\eta_f)]$. 
Consequently, the total electromagnetic power spectrum is
\begin{eqnarray}
 \frac{\partial \rho_{em}}{\partial \ln{k}} = \frac{\partial \rho(\vec{B})}{\partial \ln{k}} + \frac{\partial \rho(\vec{E})}{\partial \ln{k}} 
 = \frac{1}{\pi^2}\bigg(\frac{k^4}{a^4}\bigg)\bigg[\big|\alpha_k(\eta_f)\big|^2 + \big|\beta_k(\eta_f)\big|^2\bigg].
 \label{reheating em power spectrum}
\end{eqnarray}
It may be observed from the above equation that the comoving electromagnetic power spectrum is independent of time, which is due to 
the fact that the conformal symmetry of the electromagnetic field is restored or, equivalently, the Bogoliubov coefficients become time-independent in the 
reheating phase. Coming back to Eq.(\ref{reheating magnetic power spectrum1}), the magnetic power spectrum at time $\eta$ 
is found to depend on $\alpha_k(\eta_f)$, $\beta_k(\eta_f)$ and $\eta - \eta_f$. So, in the following, we will explicitly evaluate these quantities.

\begin{itemize}
 \item \underline{Determination of $\alpha_k(\eta_f)$ and $\beta_k(\eta_f)$}: 
 From Eqs.(\ref{continuity2}) and (\ref{reheating Bogoliubov2}), one can determine $\alpha_k(\eta_f)$ and $\beta_k(\eta_f)$ in terms of $k\eta_f$, 
 $\eta_0$ and the model parameter $q$, as follows 
 \begin{eqnarray}
  \alpha_k(\eta_f)&=&\frac{\sqrt{k/2}}{\bigg[1 - \frac{4B}{\eta_0^{2q}}\bigg(\frac{-\eta_f}{\eta_0}\bigg)^{2\epsilon q}\bigg]^{1/2}}
  \bigg\{\sqrt{-k\eta_f}\big[D_1~J_{\nu}(-k\eta_f) + D_2~J_{-\nu}(-k\eta_f)\big]\nonumber\\ 
  + i\sqrt{-k\eta_f}&\big[&D_1~J_{-1+\nu}(-k\eta_f) + D_2~J_{-1-\nu}(-k\eta_f)\big] 
  + \frac{i}{\sqrt{-k\eta_f}}\big[D_1\big(\nu - \frac{1}{2}\big)~J_{\nu}(-k\eta_f) - D_2\big(\nu + \frac{1}{2}\big)~J_{-\nu}(-k\eta_f)\big]\bigg\}\nonumber\\
 \label{reheating Bogoliubov4}
 \end{eqnarray}
 and
 \begin{eqnarray}
  \beta_k(\eta_f)&=&\frac{\sqrt{k/2}}{\bigg[1 - \frac{4B}{\eta_0^{2q}}\bigg(\frac{-\eta_f}{\eta_0}\bigg)^{2\epsilon q}\bigg]^{1/2}}
  \bigg\{\sqrt{-k\eta_f}\big[D_1~J_{\nu}(-k\eta_f) + D_2~J_{-\nu}(-k\eta_f)\big]\nonumber\\ 
  - i\sqrt{-k\eta_f}&\big[&D_1~J_{-1+\nu}(-k\eta_f) + D_2~J_{-1-\nu}(-k\eta_f)\big] 
  - \frac{i}{\sqrt{-k\eta_f}}\big[D_1\big(\nu - \frac{1}{2}\big)~J_{\nu}(-k\eta_f) - D_2\big(\nu + \frac{1}{2}\big)~J_{-\nu}(-k\eta_f)\big]\bigg\},\nonumber\\
 \label{reheating Bogoliubov5}
 \end{eqnarray}
 respectively. For the modes around the CMB mode (on which we are interested eventually, to determine the current magnetic strength), we have the relation 
 $-k\eta_f = e^{-N_f}$ with $N_f$ being the inflation e-folding number. Thereby, the Bessel function 
 $J_{\nu}(-k\eta_f)$ present in the above expression has the following asymptotic form
\begin{eqnarray}
 \lim_{|k\eta|\ll1}~J_{\nu}(-k\eta) = \frac{1}{2^{\nu}\Gamma(\nu + 1)}\big(-k\eta\big)^{\nu}\nonumber
\end{eqnarray}
and also similar asymptotic forms hold for $J_{-\nu}(-k\eta_f)$, $J_{-1-\nu}(-k\eta_f)$. Hence, it is evident from 
Eqs.(\ref{reheating Bogoliubov4}) and (\ref{reheating Bogoliubov5}) that $\alpha_k(\eta_f)$ and $\beta_k(\eta_f)$ contain terms like 
$\big(-k\eta_f\big)^{-\nu - 1/2}$ and, due to the fact that $\nu$ is positive (see Fig.[\ref{plot1}]), the presence of 
$\big(-k\eta_f\big)^{-\nu - 1/2}$ makes $\big|\alpha_k(\eta_f)\big|$, $\big|\beta_k(\eta_f)\big|$  much larger than one.

\item \underline{Determination of ``$\eta - \eta_f$'' during reheating}: The term $k(\eta - \eta_f)$ in Eq.(\ref{reheating magnetic power spectrum1}) 
actually 
leads to the non-conventional dynamics of the magnetic field. Due to the constant equation of state during reheating dynamics 
this special term boils down to the following simple form: 
\begin{eqnarray}
 k\big(\eta - \eta_f\big) = \frac{2k}{\big(3\omega_\mathrm{eff} + 1\big)}\bigg[\frac{1}{aH} - \frac{1}{a_fH_f}\bigg],
 \label{reheating conformal 1}
\end{eqnarray}
where we used $\eta - \eta_f = \int^{a}_{a_f}\frac{da}{a^2H}$. 

\end{itemize}
With the above expressions of $\alpha_k(\eta_f)$, $\beta_k(\eta_f)$ and $k(\eta - \eta_f)$, the magnetic power spectrum during the reheating 
phase in Eq.(\ref{reheating magnetic power spectrum1}) becomes \cite{Kobayashi:2019uqs}
\begin{eqnarray}
 \frac{\partial \rho(\vec{B})}{\partial \ln{k}} = \frac{1}{\pi^2}\bigg(\frac{k^4}{a^4}\bigg)\big|\beta_k(\eta_f)\big|^2~ 
 \bigg\{\mathrm{Arg}\big[\alpha_k(\eta_f)~\beta_k^{*}(\eta_f)\big] - \pi - \bigg(\frac{4k}{3\omega_\mathrm{eff} + 1}\bigg)
 \bigg(\frac{1}{aH} - \frac{1}{a_fH_f}\bigg)\bigg\}^2.
 \label{reheating magnetic power spectrum2}
\end{eqnarray}
Thereby, the evolution of the magnetic power in the reheating epoch is controlled by two different terms, namely 
the conventional one associated with the redshift factor $\propto a^{-4}$, emerging from the first term 
in the right-hand side of the Eq.(\ref{reheating magnetic power spectrum2}), and another term associated with the redshift factor 
$\propto (a^3H)^{-2}$, which has emerged out from $\frac{1}{aH}$. Since the universe expands with deceleration, 
i.e. $\omega_\mathrm{eff} > -\frac{1}{3}$, the magnetic power 
would be eventually dominated by the component $\propto (a^3H)^{-2}$. The term proportional to $(a^3H)^{-2}$ carries the main difference 
in inflationary magnetogenesis between the two cases: (i) instantaneous reheating and (ii) a reheating phase with non-zero e-fold number. 
Actually, in the context of instantaneous reheating, the magnetic power goes down as $a^{-4}$ after inflation and until today, unlike in the case 
of the reheating phase with non-zero e-fold number, where the magnetic power goes as $(a^3H)^{-2}$ from the end of inflation to the end of reheating, 
and only then as $a^{-4}$ until the present epoch. Therefore as a whole, due to the presence of the reheating phase, 
the magnetic field's present strength will be larger in comparison to that for instantaneous reheating. 
However the electric power in the reheating epoch goes by the conventional way $\propto a^{-4}$, in particular,
\begin{eqnarray}
 \frac{\partial \rho(\vec{E})}{\partial \ln{k}} = \frac{1}{\pi^2}\bigg(\frac{k^4}{a^4}\bigg)\big|\beta_k(\eta_f)\big|^2
 \label{reheating electric power spectrum2}
\end{eqnarray}
Hence the electric and magnetic power evolve differently in the reheating phase, in particular, the electric power 
goes down as $a^{-4}$ while the magnetic power  as $(a^3H)^{-2}$. 

\subsection{Current magnetic strength and constraints on $\omega_\mathrm{eff}$}
 The presence of a reheating phase with non-zero e-fold number leads to the electric field continuing to exist during post 
inflationary era until the universe becomes purely conductive and this generally happens after the end of the reheating epoch. The 
strong electric field during reheating induces the magnetic field evolution and can support the production of sufficient strength of magnetic field to survive at present time. For the Kamionkowski reheating model (considered in the present work), 
the universe is supposed to expand with some constant EoS parameter ($\omega_\mathrm{eff} > -\frac{1}{3}$ such that the expansion decelerates) in the 
reheating phase. Thus, the Hubble parameter at the end of reheating ($H_{re}$) can be related to that at the end of inflation ($H_f$) as
\begin{eqnarray}
 H_{re} = H_f\bigg(\frac{a_{re}}{a_f}\bigg)^{-\frac{3}{2}(1 + \omega_{eff})},
 \label{reheating end Hubble parameter}
\end{eqnarray}
where the suffix 're' denotes the end point of reheating and  the scale factor $a_{re}$ can be identified as 
$\frac{a_{re}}{a_f} = e^{N_\mathrm{re}}$ with $N_\mathrm{re}$ being the e-fold number of the reheating epoch and given in Eq.(\ref{reheating e-folding}). Consequently, 
from Eq.(\ref{reheating magnetic power spectrum2}) we evaluate the magnetic power spectrum at the end instant of reheating, as
\begin{eqnarray}
 \frac{\partial \rho(\vec{B})}{\partial \ln{k}}\bigg|_{re} = \frac{1}{\pi^2}\bigg(\frac{k^4}{a_{re}^4}\bigg)\big|\beta_k(\eta_f)\big|^2~ 
 \bigg\{\mathrm{Arg}\big[\alpha_k(\eta_f)~\beta_k^{*}(\eta_f)\big] - \pi - \bigg(\frac{4}{3\omega_\mathrm{eff} + 1}\bigg)
 \bigg(\frac{k}{a_fH_f}\bigg)\bigg[\bigg(\frac{H_f}{H_{re}}\bigg)^{\frac{3\omega_\mathrm{eff} + 1}{3\omega_\mathrm{eff} + 3}} - 1\bigg]\bigg\}^2~~~.
 \label{reheating magnetic power spectrum3}
\end{eqnarray}
As mentioned earlier,  after the reheating the conductivity of the universe becomes sufficiently large. In consequence, 
the electric field dies out very fast, and the magnetic field redshifts does it as $a^{-4}$ till today. The present-day magnetic power spectrum 
obeys the following relation
\begin{eqnarray}
 \frac{\partial \rho(\vec{B})}{\partial \ln{k}}\bigg|_{0} = \bigg(\frac{a_{re}}{a_0}\bigg)^4~\frac{\partial \rho(\vec{B})}{\partial \ln{k}}\bigg|_{{re}}
 \label{reheating magnetic strength 1}
\end{eqnarray}
and, as a result, Eq.(\ref{reheating magnetic power spectrum3})  leads to the current magnetic strength, as
\begin{eqnarray}
 B_0 = \frac{\sqrt{2}}{\pi}\bigg(\frac{k}{a_0}\bigg)^2\big|\beta_k(\eta_f)\big|~ 
 \bigg\{\mathrm{Arg}\big[\alpha_k(\eta_f)~\beta_k^{*}(\eta_f)\big] - \pi - \bigg(\frac{4}{3\omega_\mathrm{eff} + 1}\bigg)
 \bigg(\frac{k}{a_fH_f}\bigg)\bigg[\bigg(\frac{H_f}{H_{re}}\bigg)^{\frac{1 + 3\omega_\mathrm{eff}}{3 + 3\omega_\mathrm{eff}}} - 1\bigg]\bigg\}~~~,
 \label{reheating magnetic strength 2}
\end{eqnarray}
where we recall that the Bogoliubov coefficients have been obtained in Eqs.(\ref{reheating Bogoliubov4}) and (\ref{reheating Bogoliubov5}), respectively. 
From the above expression, we may observe that the magnetic field's present amplitude explicitly 
depends on the reheating parameters ($\omega_{eff}$ and $H_{re}$) as well as on some inflationary parameters. Thus the current magnetic strength 
encodes the information of various cosmological epochs of the universe, in particular the reheating and the inflationary epoch. As a consequence, probing 
$B_0$ opens up in turn a window for probing the early stage of the universe, particularly the reheating phase
through the current observational amplitude of the magnetic field. Actually the expression in Eq.(\ref{reheating magnetic strength 2}) shows 
a direct one-to-one correspondence between the current magnetic strength $B_0$ and the effective reheating EoS parameter $\omega_\mathrm{eff}$. 
Interestingly, the effective equation of state is no longer a free parameter as it is fixed by the $B_0$ via CMB.

Having obtained the final expression of $B_0$, now we confront our model with the CMB observations which put a constraint 
on the current magnetic strength, as $10^{-10}\mathrm{G} \lesssim B_0 \lesssim 10^{-22}\mathrm{G}$. For this purpose, we need $\frac{H_f}{H_{re}}$, which 
depends on $N_\mathrm{re}$, as discussed after Eq.(\ref{reheating end Hubble parameter}). The reheating e-fold number ($N_\mathrm{re}$) has the expression shown 
in Eq.(\ref{reheating e-folding}) and in turn requires $H_0$ and $H_f$ (i.e. the Hubble parameter at the beginning and at the end of 
inflation, respectively). We consider $H_0 = 10^{-5}M_{Pl}$, $\epsilon = 0.1$ and $N_f = 58$, 
by which $H_f$ can be estimated from Eq.(\ref{cosmic Hubble parameter}) as $H_f = H_0\exp{\bigg(\frac{-\epsilon N_f}{1+\epsilon}\bigg)} = 
5.1\times10^{-8}M_{Pl}$. Moreover, the model parameter $q$ is taken as $q = 0.5$ and recall $\eta_0^{-1} = H_0$. With such considerations, 
we plot $B_0$ versus $\omega_{eff}$ by using Eq.(\ref{reheating magnetic strength 2}) (see Fig.[\ref{plot_magnetic strength}]):

\begin{figure}[!h]
\begin{center}
\centering
\includegraphics[width=3.5in,height=2.5in]{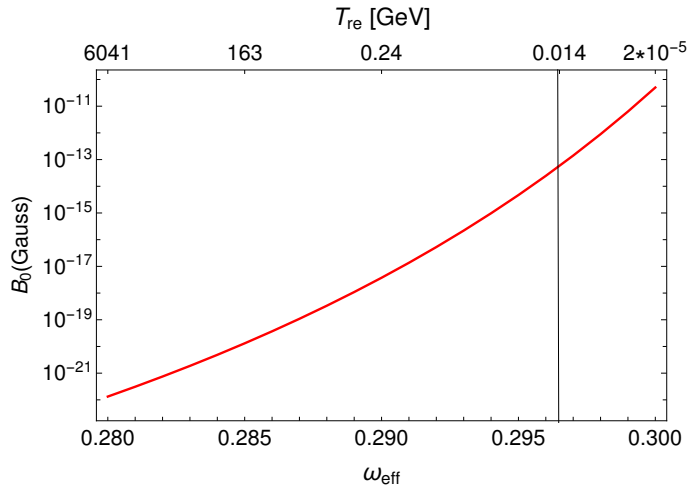}
\caption{$B_0$ (in Gauss units) vs $\omega_\mathrm{eff}$, corresponding to $\frac{k}{a_0} = 0.02\mathrm{Mpc}^{-1}$. The reheating temperature 
for different values of $\omega_\mathrm{eff}$ is shown in the upper label of the $x$-axis.}
 \label{plot_magnetic strength}
\end{center}
\end{figure}

Fig.[\ref{plot_magnetic strength}] clearly demonstrates that the theoretical prediction of $B_0$ lies well within the observational constraints 
for a certain range of values of the reheating EoS parameter, given by $0.28 \lesssim \omega_\mathrm{eff} \lesssim 0.30$. 
The upper label of the $x$-axis of Fig.[\ref{plot_magnetic strength}] shows the respective reheating temperature ($T_\mathrm{re}$) 
for different values of $\omega_\mathrm{eff}$, obtained from Eq.(\ref{reheating temperature}). Moreover, due to the BBN constraint, $T_\mathrm{re}$ is limited 
by $T_\mathrm{re} \gtrsim 10^{-2}\mathrm{GeV}$, which is shown by the vertical line in Fig.[\ref{plot_magnetic strength}]. Such 
constraint on $T_\mathrm{re}$, in turn, sets a lower permissible limit on $\omega_\mathrm{eff}$, in particular $\omega_\mathrm{eff} \gtrsim 0.2965$, as shown in 
the same figure.  Therefore, in order to make the present model compatible both with 
the CMB observations on $B_0$ and the BBN constraint on $T_\mathrm{re}$, the viable 
regime of $\omega_\mathrm{eff}$ turns out to be: $0.28 \lesssim \omega_\mathrm{eff} \lesssim 0.2965$.

Therefore, the presence of the reheating phase with a non-zero e-fold number does in fact enhance the strength of 
the magnetic fields surviving in the present epoch and makes the theoretical predictions of the model fully compatible 
with the most recent observations. This is to be compared to the usual instantaneous reheating case, where the magnetic 
field's current strength is always found to be far below  the present observational constraints.

\section{Constraints from Schwinger backreaction}\label{sec_schwinger}

In the earlier section, we consider the universe to be a bad conductor during the reheating phase, which means that the EM field 
during the reheating obeys the standard Maxwell's equations in vacuum. However, due to the Schwinger production, 
the assumption of zero conductivity demands 
a proper investigation, which is the subject of the present section.

Eqs.(\ref{reheating magnetic power spectrum2}) and 
(\ref{reheating electric power spectrum2}) clearly indicate that the ratio of magnetic to electric power spectrum goes by,
\begin{eqnarray}
 \frac{\mathcal{P}_{B}}{\mathcal{P}_{E}} = 
 \bigg\{\mathrm{Arg}\big[\alpha_k(\eta_f)~\beta_k^{*}(\eta_f)\big] - \pi - \bigg(\frac{4k}{3\omega_\mathrm{eff} + 1}\bigg)
 \bigg(\frac{1}{aH} - \frac{1}{a_fH_f}\bigg)\bigg\}^2
 \label{schwinger1}
\end{eqnarray}
with $\mathcal{P}_B = \frac{\partial\rho(B)}{\partial\ln{k}}$, $\mathcal{P}_E = \frac{\partial\rho(E)}{\partial\ln{k}}$. Moreover, recall, 
the Bogoliubov coefficients at $\eta_f$ are given by,
\begin{eqnarray}
 \alpha_k(\eta_f)&=&\sqrt{\frac{k}{2}}~A(k,\eta_f) + \frac{i}{\sqrt{2k}}~A'(k,\eta_f),\nonumber\\
 \beta_k(\eta_f)&=&\sqrt{\frac{k}{2}}~A(k,\eta_f) - \frac{i}{\sqrt{2k}}~A'(k,\eta_f)~~~~.
 \label{schwinger2}
\end{eqnarray}
with $A(k,\eta_f)$ is shown in Eq.(\ref{rehearting continuity3}). The presence of $J_{-\nu}(-k\eta)$ in the superhorizon solution 
of the EM mode function makes 
$\frac{dA(k,\eta)}{d(-k\eta)}\big|_{\eta_f}$ much larger than $A(k,\eta_f)$. Consequently the Bogoliubov coefficients from Eq.(\ref{schwinger2}) can be 
expressed as,
\begin{eqnarray}
 \alpha_k(\eta_f) = \frac{i}{\sqrt{2k}}~A'(k,\eta_f)~~~~,~~~~~~\beta_k(\eta_f) = -\frac{i}{\sqrt{2k}}~A'(k,\eta_f)~~~~,
 \label{schwinger3}
\end{eqnarray}
which further leads to the relative phase factors of the Bogoliubov coefficients at $\eta = \eta_f$ as 
$\mathrm{Arg}\big[\alpha_k(\eta_f)~\beta_k^{*}(\eta_f)\big] \simeq \pi$. As a result, the ratio $\frac{\mathcal{P}_{B}}{\mathcal{P}_{E}}$ from 
Eq.(\ref{schwinger1}) takes the following form,
\begin{eqnarray}
 \frac{\mathcal{P}_{B}}{\mathcal{P}_{E}} = \bigg\{\bigg(\frac{4k}{3\omega_\mathrm{eff} + 1}\bigg)
 \bigg(\frac{1}{aH} - \frac{1}{a_fH_f}\bigg)\bigg\}^2~~.
 \label{schwinger4}
\end{eqnarray}
Here it may be mentioned that we are interested in determining the magnetic strength around the CMB scales which, in fact, lies within the superhorizon 
regime in the reheating phase, in particular the mode $k$ in the above expression satisfies $k < aH$. Hence the above equation implies that the electric 
power spectrum is stronger compared to the magnetic power spectrum in the reheating epoch. Such a strong electric field can give rise to 
Schwinger production of charged particles, which in turn may backreact the magnetogenesis scenario \cite{Kobayashi:2019uqs,Kobayashi:2014zza,
Stahl:2018idd,Rajeev:2019okd}. 
Moreover the Authors of \cite{Kobayashi:2019uqs} themselves acknowledged in their paper that 
it is not at all obvious that the conductivity is negligible during preheating; indeed it might be so for a certain time 
but eventually it may catch up with the enormous values we see in the thermalised plasma. Therefore it is important to discuss 
the Schwinger backreaction in the present context by considering a non-zero electrical conductivity of the universe during the reheating phase.

\begin{figure}[!h]
\begin{center}
\centering
\includegraphics[width=4.0in,height=2.5in]{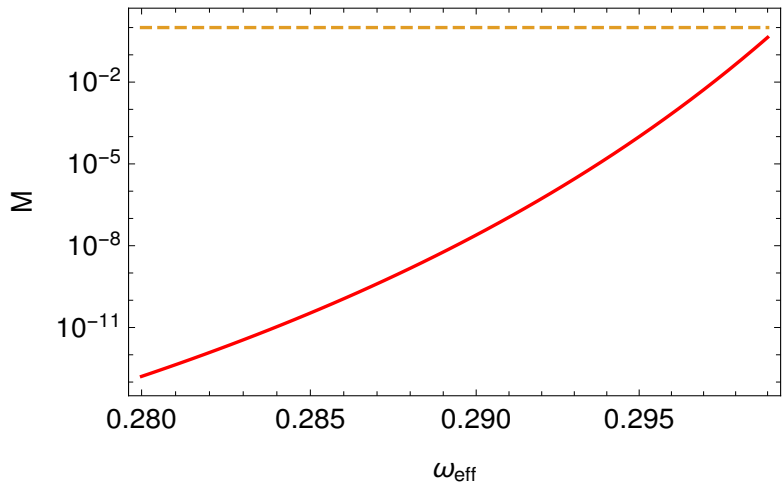}
\caption{$\mathcal{M}$ vs $\omega_\mathrm{eff}$ by the solid curve, corresponding to $\frac{k}{a_0} = 0.02\mathrm{Mpc}^{-1}$. 
Moreover we consider $g_{re} = g_{s,re} = 100$ and $T_0 = 2.3\times10^{-13}\mathrm{GeV}$. The dashed horizontal curve denotes the conatant value $= 1$.}
 \label{plot_schwinger}
\end{center}
\end{figure}

In presence of non-zero conductivity, symbolized by $\sigma$, the Ampere-Maxwell equation of the EM field reads as,
\begin{eqnarray}
 \epsilon_{ijl}\partial_jB_l = E_i' + \bigg(a\sigma + \frac{a'}{a}\bigg)E_i
 \label{sch1}
\end{eqnarray}
where $\epsilon_{ijl}$ is the 3-Levi Civita symbol. Thereby the condition under which the Schwinger backreaction can be neglected is given by,
\begin{eqnarray}
 \big|a\sigma| < \frac{a'}{a} = aH~~~~~~~~~\mathrm{or},~~~~~~~~~~\big|\sigma| < H~~.
 \label{sch2}
\end{eqnarray}
To progress further, we need a certain form of $\sigma$ in the FRW cosmological background where the Hubble parameter 
evolves as $H \propto a^{-\frac{3}{2}\big(1 + \omega_\mathrm{eff}\big)}$. A complete analysis of $\sigma$ in a generic FRW spacetime is beyond 
the scope of this paper. However the electrical conductivity in FRW background can be estimated from that of in the Minkowski background by replacing 
the elapsed time of the electric field in the Minkowski spacetime to $H^{-1}$ in FRW spacetime, as explained in \cite{Kobayashi:2019uqs}. 
This results the conductivity in FRW spacetime as,
\begin{eqnarray}
 \sigma = \frac{1}{12\pi^3}\frac{|e^3E|}{H}\exp{\big[-\frac{\pi m^2}{|eE|}\big]}
 \label{sch3}
\end{eqnarray}
where $m$ and $e$ are the mass and the electrical charge of the produced charged particles respectively. Here we would like to mention that the 
above form of $\sigma$ is also valid for a de-Sitter background spacetime under the strong electric field approximation \cite{Kobayashi:2014zza}. 
Eq.(\ref{sch3}) can be solved for the electric field as,
\begin{eqnarray}
 \frac{|E|}{H^2} = \frac{12\pi^3}{e^3}\bigg(\frac{\sigma}{H}\bigg)\exp{\bigg\{W\bigg(\frac{e^2}{12\pi^2}\frac{m^2}{H^2}\frac{H}{\sigma}\bigg)\bigg\}}
 \label{sch4}
\end{eqnarray}
where $W(x)$ is known as the Lambert W-function and corresponds to the solution of $We^{W} = x$. For low massive charged particles i.e for 
$m^2 \ll H^2$, the W-function can be approximated as $W(0 < x \ll 1) \approx x$ and thus the above equation can be expressed as,
\begin{eqnarray}
 \frac{|E|}{H^2} = \frac{12\pi^3}{e^3}\bigg(\frac{\sigma}{H}\bigg)~~.
 \label{sch5}
\end{eqnarray}
Thereby the above expression translates the condition of Eq.(\ref{sch2}) to an upper bound of the electric field during the reheating phase as,
\begin{eqnarray}
 \frac{|E|}{H^2} < \frac{12\pi^3}{e^3} \approx 12\pi^3,
 \label{sch6}
\end{eqnarray}
where the electric field ($E$) during the reheating phase depends on the Bogoliubov coefficient $\beta_k(\eta_f)$, as indicated by 
Eq.(\ref{reheating electric power spectrum2}). The superhorizon solution of the EM mode function (see Eq.(\ref{rehearting continuity3})) immediately 
leads to the following expression of $\beta_k(\eta_f)$,
\begin{eqnarray}
 \big|\beta_k(\eta_f)\big| = \bigg|\frac{2B\epsilon q}{\eta_0^{2q}\big(\frac{2B}{\eta_0^{2q}} - 1\big)}\bigg|\big(-k\eta_f\big)^{-\nu-\frac{1}{2}}
 \label{sch7}
\end{eqnarray}
where we use $\nu^2 = \frac{1}{4} - \frac{4B\epsilon q}{\eta_0^{2q}\big(2B/\eta_0^{2q} - 1\big)}$ with 
$B = \kappa^{2q}\big\{\big[6\beta(\beta + 1)\big]^q + \big[-24(\beta + 1)^3\big]^{q/2}\big\}$ and, recall, $q$ is the model parameter 
which is generally taken as $q = 0.5$ during the determination of the present magnetic strength in Sec.[\ref{sec_kamionkowski}]. Moreover, as mentioned 
earlier, the reheating phase is dominated by the constant EoS $\omega_\mathrm{eff}$ and thus the Hubble 
parameter in Eq.(\ref{sch6}) follows the evolution as,
\begin{eqnarray}
 H^2 = H_f^2\bigg(\frac{a}{a_f}\bigg)^{-3\big(1 + \omega_\mathrm{eff}\big)}
 \label{sch8}
\end{eqnarray}
with $H_f$ being the Hubble parameter at the end of inflation and given by $H_f = H_0\exp{\big[-\frac{\epsilon N_f}{1 + \epsilon}\big]}$. Using the 
above expressions of $\beta_k(\eta_f)$ and $H^2$, the inequality of Eq.(\ref{sch6}) turns out to be,
\begin{eqnarray}
 \frac{1}{12\pi^3}\bigg|\frac{2B\epsilon q}{\eta_0^{2q}\big(\frac{2B}{\eta_0^{2q}} - 1\big)}\bigg|\big(-k\eta_f\big)^{\frac{3}{2} - \nu}
 \bigg(\frac{a}{a_f}\bigg)^{1 + 3\omega_\mathrm{eff}} < 1~~.
 \label{sch9}
\end{eqnarray}
As evident, the quantity in the left hand side of the above inequality 
increases with the scale factor during the reheating phase and thus it attains the maximum value at the end of reheating i.e at $a = a_\mathrm{re}$. 
Thereby we define,
\begin{eqnarray}
 \mathcal{M} = \frac{1}{12\pi^3}\bigg|\frac{2B\epsilon q}{\eta_0^{2q}\big(\frac{2B}{\eta_0^{2q}} - 1\big)}\bigg|\big(-k\eta_f\big)^{\frac{3}{2} - \nu}
 \bigg(\frac{a_\mathrm{re}}{a_f}\bigg)^{1 + 3\omega_\mathrm{eff}}
 \label{sch10}
\end{eqnarray}
which is the maximum value of the left hand side quantity of Eq.(\ref{sch9}). The factors $k\eta_f$ and $\frac{a_\mathrm{re}}{a_f}$ present 
in the above expressions are connected with the e-folding number of inflation and reheating era respectively, in particular, they are given by: 
$-k\eta_f = e^{-N_f}$ and $\frac{a_\mathrm{re}}{a_f} = e^{N_\mathrm{re}}$, where $k$ is considered to be around the CMB scale and 
the $N_\mathrm{re}$ (in terms of reheating EoS and $N_f$) is shown 
in Eq.(\ref{reheating e-folding}). Plugging back such forms of $k\eta_f$ and $\frac{a_\mathrm{re}}{a_f}$ into 
Eq.(\ref{sch10}), we get
\begin{eqnarray}
 \mathcal{M} = \bigg(\frac{\epsilon q\big(12^q + 24^{q/2}\big)}{6\pi^3}\bigg)\bigg(\frac{H_0}{M_{Pl}}\bigg)^{2q}~
 \exp{\bigg\{-\bigg(\frac{3}{2}-\nu\bigg)N_f + \big(1 + 3\omega_\mathrm{eff}\big)N_\mathrm{re}\bigg\}}
 \label{sch11}
\end{eqnarray}
where we use the form of $B$ (as shown earlier after Eq.(\ref{sch7})). With the expression of $N_\mathrm{re}$ in Eq.(\ref{reheating e-folding}), 
we determine the argument within the exponential term as follows,
\begin{eqnarray}
 -\bigg(\frac{3}{2}-\nu\bigg)N_f&+&\big(1 + 3\omega_\mathrm{eff}\big)N_\mathrm{re} = 
 -N_f\bigg[\bigg(\frac{3}{2} - \nu\bigg) + \frac{4\big(1+3\omega_\mathrm{eff}\big)}{\big(1 - 3\omega_\mathrm{eff}\big)}\bigg]\nonumber\\ 
 &-&\frac{4\big(1 + 3\omega_\mathrm{eff}\big)}{\big(1 - 3\omega_\mathrm{eff}\big)}
 \bigg[\frac{1}{4}\ln{\bigg(\frac{45}{\pi^2g_{re}}\bigg)} + \frac{1}{3}\ln{\bigg(\frac{11g_{s,re}}{43}\bigg)} 
 + \ln{\bigg(\frac{k}{a_0T_0}\bigg)} + \ln{\bigg(\frac{(3H_f^2M_\mathrm{Pl}^2)^{1/4}}{H_0}\bigg)}\bigg]~~.
 \label{sch12}
\end{eqnarray}
Thereby Eqs.(\ref{sch11}) and (\ref{sch12}) clearly indicate that $\mathcal{M}$ depends on some 
inflationary parameters like $H_0/M_{Pl}$, $\epsilon$, $N_f$; the reheating EoS i.e $\omega_\mathrm{eff}$; and the model parameter $q$. Recall, 
the parametric regime that we have considered in determining the current magnetic strength ($B_0$) in the previous section are given by: 
$H_0 = 10^{-5}M_{Pl}$, $\epsilon = 0.1$, $N_f = 58$ and $q = 0.5$ respectively, for which the Hubble parameter at the end of inflation 
comes as $H_f = H_0\exp{\big[\frac{-\epsilon N_f}{1+\epsilon}\big]} = 5.1\times10^{-8}M_{Pl}$. With this same parametric space and by using 
Eq.(\ref{sch11}), here we give the plot of $\mathcal{M}$ versus $\omega_\mathrm{eff}$ by the solid curve in Fig.[\ref{plot_schwinger}], while 
the dashed horizontal curve in the figure corresponds to the constant value $=1$.\\

Fig.[\ref{plot_schwinger}] clearly demonstrates that the quantity $\mathcal{M}$ lies below unity for $\omega_\mathrm{eff} \lesssim 0.299$, which in turn 
confirms the inequality of Eq.(\ref{sch9}) during the entire reheating epoch. Thereby we may argue that 
in the regime $\omega_\mathrm{eff} \lesssim 0.299$, the Schwinger backreaction or equivalently the electrical conductivity in the reheating phase 
may be neglected and consequently the evolution equation of the electromagnetic field can be regarded as the standard Maxwell's equation in vacuum. 
Here we need to recall from Fig.[\ref{plot_magnetic strength}] that in order to make the magnetic field's current strength ($B_0$) 
compatible with the CMB observations, the viable regime of $\omega_\mathrm{eff}$ is found to be: $0.28 \lesssim \omega_\mathrm{eff} \lesssim 0.2965$ 
which is, in fact, a sub-part of $\omega_\mathrm{eff} \lesssim 0.299$. Thereby the regime of the reheating EoS that makes the model viable in regard to the 
CMB observations on $B_0$ as well as leads to a negligible Schwinger backreaction is given by $0.28 \lesssim \omega_\mathrm{eff} \lesssim 0.2965$.\\

Before concluding, it deserves mentioning that the cosmological models characterized by vectors non-minimally coupled to curvature, in particular 
the action that contain an effective potential term for the vector field (i.e $V(A_{\mu}A^{\mu})$) may lead to ghost in the model. The presence of 
$V(A^2)$ in the action spoils the U(1) invariance of the electromagnetic field, due to which the massive vector field gets three physical degrees 
of freedom: two of them are usual transverse modes and the other one is the longitudinal mode. The longitudinal mode with momentum $p^2 > |M^2|$ 
(where $M^2$ is the effective mass squared of the vector field) appears as a ghost field in the model, i.e a field with negative kinetic energy, 
whenever $M^2 < 0$ \cite{Himmetoglu:2009qi,Himmetoglu:2008zp,Himmetoglu:2008hx,Karciauskas:2010as}. 
One of such example where the vector field has a negative $M^2$, is given by the following action \cite{Turner:1987bw}:
\begin{eqnarray}
 S = \int d^4x\sqrt{-g}\bigg[-\frac{1}{4}F_{\mu\nu}F^{\mu\nu} + \frac{1}{12}RA^2\bigg]~~,
 \label{new action}
\end{eqnarray}
where $V(A^2) = -\frac{1}{12}RA^2$ and thus the effective mass squared of the vector field comes as 
$M^2 = -R/6$ which is indeed negative.\\ 
However on contrary, in the present work, it is the $kinetic~term$ of the electromagnetic field that 
gets coupled with the spacetime curvature, in particular the EM Lagrangian is $\mathcal{L} \sim f(R,\mathcal{G})F_{\mu\nu}F^{\mu\nu}$ 
(see Eq.(\ref{actuon part 3})), and hence there is no potential term of the electromagnetic field appearing in the action. Thus the 
U(1) invariance is preserved in the present model, and consequently the problematic longitudinal mode is absent. 
Thereby we may argue that in the present model 
where the kinetic term of the EM field couples with the background Ricci scalar and the Gauss-Bonnet curvature, is free from ghost fields. However such 
kind of argument needs a study of full perturbation analysis, which is beyond the scope of this paper and thus expected to study in future.

\section{Conclusion}
We have constructed a viable inflationary magnetogenesis model where the electromagnetic field couples with the spacetime curvature, specifically, with the Ricci scalar and the Gauss-Bonnet invariant. The background spacetime is controlled by the well-studied, and mathematically well-grounded, scalar-Einstein-Gauss-Bonnet gravity theory, which is known to provide a good inflationary model for suitable choices of the Gauss-Bonnet coupling function and the scalar field potential. The model has some quite remarkable features. First, the non-minimal coupling between the electromagnetic field and the curvature breaks down the conformal invariance of the  field and, therefore, it does not require any additional coupling to the scalar field. A second key feature is that, as the curvature is significant in the
early universe, the conformal breaking term ($S_{CB}$) introduces a non-trivial correction to the electromagnetic action; however, at
late times (in particular after the end of inflation), $S_{CB}$ does not actually  contribute, and the electromagnetic field 
behaves according to the standard Maxwell equations. Thirdly, the conformal breaking coupling is suppressed by $\kappa^{2q}$ 
(with $q$ being the model parameter) and, thus, it does not lead to the strong coupling problem for $q \sim \mathcal{O}(1)$, which affects so many models, as this value also moves within a viable parametric regime, in regard to the compatibility of the magnetogenesis model with the most reliable observational data. As last strong point of our model, the electromagnetic field is found to have a negligible backreaction on the background spacetime, and thus the backreaction issue in the present magnetogenesis model is naturally resolved.

In such scenario, we have explored the evolution of the electric and magnetic fields with the expansion of the universe, starting from the inflationary 
era. During the cosmic evolution, the universe enters into a reheating phase after the inflation epoch, and depending on the reheating mechanism, we have considered 
two different cases: (i) as the first one, we assumed an instantaneous reheating where the universe makes a sudden jump from the inflationary epoch to a radiation dominated stage, during which the cosmic Hubble parameter goes as 
$H \propto t^{-1}$, with $t$ being the cosmic time; and (ii) a case where the universe experiences a reheating phase, with non-zero e-fold number; 
in particular, we have considered the conventional reheating mechanism proposed by Kamionkowski et al. \cite{Dai:2014jja}, where the main idea is to parametrize the 
reheating phase by a constant effective equation of state parameter ($\omega_\mathrm{eff}$).

We have shown above that, in the instantaneous reheating case, the conductivity becomes large immediately after inflation and, consequently, the electric field 
dies out quite rapidly. This, along with the fact that the electromagnetic field acquires conformal symmetry after inflation, leads to the evolution 
of the magnetic field energy density to go as $a^{-4}$ during the post inflationary epoch. As a result, the theoretical predictions for the  present amplitude of the magnetic field ($B_0$) was found to lie in fact far below  the observational constraint given by  $10^{-22}\mathrm{G} \lesssim B_0 \lesssim 10^{-10}\mathrm{G}$.

However, we have seen that the scenario becomes completely different when the reheating phase is considered to have a non-zero e-fold number. In this case, the 
conductivity remains non-zero and, consequently, the strong electric field induces a magnetic field evolution during the epoch between 
the end of inflation and until the end of reheating. On the other hand, after the reheating phase, the universe becomes a good conductor and, 
hence, the electric field goes to zero. Specifically, the magnetic field energy density evolves as $(a^3H)^{-2}$ during the reheating era, and 
later, as $a^{-4}$, as usual (this means, from the end of the reheating stage up to the present epoch). Such evolution clearly indicates that the presence of a reheating phase with non-vanishing e-fold number enhances the amplitude of the magnetic field's current, as compared to ordinary
instantaneous reheating, where the magnetic field energy density decays as $a^{-4}$ from the very end of inflation. 

As a consequence, the 
present strength of the magnetic field falls within the range dictated by the observational constraints, for a suitable regime of reheating parameters, thus overcoming the severe problems of the instantaneous 
reheating case. Moreover, the evolution of the magnetic field through the reheating phase allows to encode very valuable information about the
reheating EoS parameter ($\omega_\mathrm{eff}$) and about some inflationary parameters on the current magnetic field strength ($B_0$). Therefore, probing $B_0$ opens up, in turn, a window for probing the early stage of the universe, in particular, the reheating phase, through 
the current observational amplitude of the magnetic field. It has been shown in the paper that, in order to make compatible the present magnetogenesis model  both with 
the CMB observations on $B_0$ and with the BBN constraint on the reheating temperature, the viable regime 
of $\omega_\mathrm{eff}$ has to be the following: $0.28 \lesssim \omega_\mathrm{eff} \lesssim 0.2965$. This provides a viable constraint on the reheating EoS parameter from CMB observations.

%%%%%%%%%%%%%%%%%%%%%%%%%%%%%%%%%%%%%%%%%%%%%%%%%%%%%%%%%%%%%%%%%%%%%%%%%%%%
\subsection*{Acknowledgments}
The work of KB was partially supported by the JSPS KAKENHI Grant Number JP
25800136 and Competitive Research Funds for Fukushima University Faculty
(19RI017). TP acknowledges D. Maity for useful discussions.
%%%%%%%%%%%%%%%%%%%%%%%%%%%%%%%%%%%%%%%%%%%%%%%%%%%%%%%%%%%%%%%%%%%%%%%%%%%%


\begin{thebibliography}{99}
 %\cite{Grasso:2000wj}
\bibitem{Grasso:2000wj}
D.~Grasso and H.~R.~Rubinstein,
%``Magnetic fields in the early universe,''
Phys. Rept. \textbf{348} (2001), 163-266
%doi:10.1016/S0370-1573(00)00110-1
[arXiv:astro-ph/0009061 [astro-ph]].
%792 citations counted in INSPIRE as of 10 Dec 2020

%\cite{Beck:2000dc}
\bibitem{Beck:2000dc}
R.~Beck,
%``Galactic and extragalactic magnetic fields,''
Space Sci. Rev. \textbf{99} (2001), 243-260
%doi:10.1023/A:1013805401252
[arXiv:astro-ph/0012402 [astro-ph]].
%202 citations counted in INSPIRE as of 10 Dec 2020


%\cite{Widrow:2002ud}
\bibitem{Widrow:2002ud}
L.~M.~Widrow,
%``Origin of galactic and extragalactic magnetic fields,''
Rev. Mod. Phys. \textbf{74} (2002), 775-823
%doi:10.1103/RevModPhys.74.775
[arXiv:astro-ph/0207240 [astro-ph]].
%543 citations counted in INSPIRE as of 10 Dec 2020


%\cite{Kandus:2010nw}
\bibitem{Kandus:2010nw}
A.~Kandus, K.~E.~Kunze and C.~G.~Tsagas,
%``Primordial magnetogenesis,''
Phys. Rept. \textbf{505} (2011), 1-58
%doi:10.1016/j.physrep.2011.03.001
[arXiv:1007.3891 [astro-ph.CO]].
%260 citations counted in INSPIRE as of 10 Dec 2020


%\cite{Durrer:2013pga}
\bibitem{Durrer:2013pga}
R.~Durrer and A.~Neronov,
%``Cosmological Magnetic Fields: Their Generation, Evolution and Observation,''
Astron. Astrophys. Rev. \textbf{21} (2013), 62
%doi:10.1007/s00159-013-0062-7
[arXiv:1303.7121 [astro-ph.CO]].
%405 citations counted in INSPIRE as of 10 Dec 2020


%\cite{Subramanian:2015lua}
\bibitem{Subramanian:2015lua}
K.~Subramanian,
%``The origin, evolution and signatures of primordial magnetic fields,''
Rept. Prog. Phys. \textbf{79} (2016) no.7, 076901
%doi:10.1088/0034-4885/79/7/076901
[arXiv:1504.02311 [astro-ph.CO]].
%199 citations counted in INSPIRE as of 10 Dec 2020

%\cite{Kulsrud:2007an}
\bibitem{Kulsrud:2007an}
R.~M.~Kulsrud and E.~G.~Zweibel,
%``The Origin of Astrophysical Magnetic Fields,''
Rept. Prog. Phys. \textbf{71} (2008), 0046091
%doi:10.1088/0034-4885/71/4/046901
[arXiv:0707.2783 [astro-ph]].
%112 citations counted in INSPIRE as of 10 Dec 2020

%\cite{Brandenburg:2004jv}
\bibitem{Brandenburg:2004jv}
A.~Brandenburg and K.~Subramanian,
%``Astrophysical magnetic fields and nonlinear dynamo theory,''
Phys. Rept. \textbf{417} (2005), 1-209
%doi:10.1016/j.physrep.2005.06.005
[arXiv:astro-ph/0405052 [astro-ph]].
%539 citations counted in INSPIRE as of 10 Dec 2020


%\cite{Subramanian:2009fu}
\bibitem{Subramanian:2009fu}
K.~Subramanian,
%``Magnetic fields in the early universe,''
Astron. Nachr. \textbf{331} (2010), 110-120
%doi:10.1002/asna.200911312
[arXiv:0911.4771 [astro-ph.CO]].
%95 citations counted in INSPIRE as of 10 Dec 2020

%\cite{Sharma:2017eps}
\bibitem{Sharma:2017eps}
R.~Sharma, S.~Jagannathan, T.~R.~Seshadri and K.~Subramanian,
%``Challenges in Inflationary Magnetogenesis: Constraints from Strong Coupling, Backreaction and the Schwinger Effect,''
Phys. Rev. D \textbf{96} (2017) no.8, 083511
%doi:10.1103/PhysRevD.96.083511
[arXiv:1708.08119 [astro-ph.CO]].
%32 citations counted in INSPIRE as of 10 Dec 2020

%\cite{Sharma:2018kgs}
\bibitem{Sharma:2018kgs}
R.~Sharma, K.~Subramanian and T.~R.~Seshadri,
%``Generation of helical magnetic field in a viable scenario of inflationary magnetogenesis,''
Phys. Rev. D \textbf{97} (2018) no.8, 083503
%doi:10.1103/PhysRevD.97.083503
[arXiv:1802.04847 [astro-ph.CO]].
%22 citations counted in INSPIRE as of 10 Dec 2020


%\cite{Jain:2012ga}
\bibitem{Jain:2012ga}
R.~K.~Jain and M.~S.~Sloth,
%``Consistency relation for cosmic magnetic fields,''
Phys. Rev. D \textbf{86} (2012), 123528
%doi:10.1103/PhysRevD.86.123528
[arXiv:1207.4187 [astro-ph.CO]].
%42 citations counted in INSPIRE as of 10 Dec 2020



%\cite{Durrer:2010mq}
\bibitem{Durrer:2010mq}
R.~Durrer, L.~Hollenstein and R.~K.~Jain,
%``Can slow roll inflation induce relevant helical magnetic fields?,''
JCAP \textbf{03} (2011), 037
%doi:10.1088/1475-7516/2011/03/037
[arXiv:1005.5322 [astro-ph.CO]].
%109 citations counted in INSPIRE as of 10 Dec 2020


%\cite{Kanno:2009ei}
\bibitem{Kanno:2009ei}
S.~Kanno, J.~Soda and M.~a.~Watanabe,
%``Cosmological Magnetic Fields from Inflation and Backreaction,''
JCAP \textbf{12} (2009), 009
%doi:10.1088/1475-7516/2009/12/009
[arXiv:0908.3509 [astro-ph.CO]].
%161 citations counted in INSPIRE as of 10 Dec 2020


%\cite{Campanelli:2008kh}
\bibitem{Campanelli:2008kh}
L.~Campanelli,
%``Helical Magnetic Fields from Inflation,''
Int. J. Mod. Phys. D \textbf{18} (2009), 1395-1411
%doi:10.1142/S0218271809015175
[arXiv:0805.0575 [astro-ph]].
%67 citations counted in INSPIRE as of 10 Dec 2020




%\cite{Demozzi:2009fu}
\bibitem{Demozzi:2009fu}
V.~Demozzi, V.~Mukhanov and H.~Rubinstein,
%``Magnetic fields from inflation?,''
JCAP \textbf{08} (2009), 025
%doi:10.1088/1475-7516/2009/08/025
[arXiv:0907.1030 [astro-ph.CO]].
%234 citations counted in INSPIRE as of 10 Dec 2020


\bibitem{Bamba:2008ja}
  K.~Bamba and S.~D.~Odintsov,
  %``Inflation and late-time cosmic acceleration in non-minimal Maxwell-$F(R)$ gravity and the generation of large-scale magnetic fields,''
  JCAP {\bf 0804} (2008) 024
  %doi:10.1088/1475-7516/2008/04/024
  [arXiv:0801.0954 [astro-ph]].
  
  
  %\cite{Bamba:2008xa}
\bibitem{Bamba:2008xa}
   K.~Bamba, S.~Nojiri and S.~D.~Odintsov,
   %``Inflationary cosmology and the late-time accelerated expansion of
   %the universe in non-minimal Yang-Mills-F(R) gravity and non-minimal
   %vector-F(R) gravity,''
   Phys.\ Rev.\ D {\bf 77}, 123532 (2008)
%  doi:10.1103/PhysRevD.77.123532
   [arXiv:0803.3384 [hep-th]].
   %%CITATION = doi:10.1103/PhysRevD.77.123532;%%
   
   
   %\cite{Bamba:2012mi}
\bibitem{Bamba:2012mi}
   K.~Bamba, C.~Q.~Geng and L.~W.~Luo,
   %``Generation of large-scale magnetic fields from inflation
   %in teleparallelism,''
   JCAP {\bf 1210}, 058 (2012)
%  doi:10.1088/1475-7516/2012/10/058
   [arXiv:1208.0665 [astro-ph.CO]].
   %%CITATION = doi:10.1088/1475-7516/2012/10/058;%%


%\cite{Bamba:2006ga}
\bibitem{Bamba:2006ga}
K.~Bamba and M.~Sasaki,
%``Large-scale magnetic fields in the inflationary universe,''
JCAP \textbf{02} (2007), 030
%doi:10.1088/1475-7516/2007/02/030
[arXiv:astro-ph/0611701 [astro-ph]].
%123 citations counted in INSPIRE as of 10 Dec 2020

%\cite{Bamba:2003av}
\bibitem{Bamba:2003av}
K.~Bamba and J.~Yokoyama,
%``Large scale magnetic fields from inflation in dilaton electromagnetism,''
Phys. Rev. D \textbf{69} (2004), 043507
%doi:10.1103/PhysRevD.69.043507
[arXiv:astro-ph/0310824 [astro-ph]].
%163 citations counted in INSPIRE as of 11 Dec 2020


%\cite{Bamba:2004cu}
\bibitem{Bamba:2004cu}
K.~Bamba and J.~Yokoyama,
%``Large-scale magnetic fields from dilaton inflation in noncommutative spacetime,''
Phys. Rev. D \textbf{70} (2004), 083508
%doi:10.1103/PhysRevD.70.083508
[arXiv:hep-ph/0409237 [hep-ph]].
%89 citations counted in INSPIRE as of 11 Dec 2020



%\cite{Kobayashi:2019uqs}
\bibitem{Kobayashi:2019uqs}
T.~Kobayashi and M.~S.~Sloth,
%``Early Cosmological Evolution of Primordial Electromagnetic Fields,''
Phys. Rev. D \textbf{100} (2019) no.2, 023524
%doi:10.1103/PhysRevD.100.023524
[arXiv:1903.02561 [astro-ph.CO]].
%20 citations counted in INSPIRE as of 10 Dec 2020


\bibitem{Bamba:2008my} 
  K.~Bamba, N.~Ohta and S.~Tsujikawa,
  %``Generic estimates for magnetic fields generated during inflation including Dirac-Born-Infeld theories,''
  Phys.\ Rev.\ D {\bf 78}, 043524 (2008)
  doi:10.1103/PhysRevD.78.043524
  [arXiv:0805.3862 [astro-ph]].
  
  
  %\cite{Giovannini:2017rbc}
\bibitem{Giovannini:2017rbc}
M.~Giovannini,
%``Probing large-scale magnetism with the Cosmic Microwave Background,''
Class. Quant. Grav. \textbf{35} (2018) no.8, 084003
doi:10.1088/1361-6382/aab17d
[arXiv:1712.07598 [astro-ph.CO]].
%9 citations counted in INSPIRE as of 08 Feb 2021


%\cite{Giovannini:2003yn}
\bibitem{Giovannini:2003yn}
M.~Giovannini,
%``The Magnetized universe,''
Int. J. Mod. Phys. D \textbf{13} (2004), 391-502
doi:10.1142/S0218271804004530
[arXiv:astro-ph/0312614 [astro-ph]].
%325 citations counted in INSPIRE as of 08 Feb 2021



%\cite{Lambiase:2004zb}
\bibitem{Lambiase:2004zb}
G.~Lambiase and A.~R.~Prasanna,
%``Gauge invariant wave equations in curved space-times and primordial magnetic fields,''
Phys. Rev. D \textbf{70} (2004), 063502
doi:10.1103/PhysRevD.70.063502
[arXiv:gr-qc/0407071 [gr-qc]].
%55 citations counted in INSPIRE as of 08 Feb 2021



%\cite{Lambiase:2008zz}
\bibitem{Lambiase:2008zz}
G.~Lambiase, S.~Mohanty and G.~Scarpetta,
%``Magnetic field amplification in f(R) theories of gravity,''
JCAP \textbf{07} (2008), 019
doi:10.1088/1475-7516/2008/07/019
%28 citations counted in INSPIRE as of 08 Feb 2021






%\cite{Ratra:1991bn}
\bibitem{Ratra:1991bn}
B.~Ratra,
%``Cosmological 'seed' magnetic field from inflation,''
Astrophys. J. Lett. \textbf{391} (1992), L1-L4
%oi:10.1086/186384
%652 citations counted in INSPIRE as of 10 Dec 2020

%\cite{Ade:2015cva}
\bibitem{Ade:2015cva}
P.~A.~R.~Ade \textit{et al.} [Planck],
%``Planck 2015 results. XIX. Constraints on primordial magnetic fields,''
Astron. Astrophys. \textbf{594} (2016), A19
%doi:10.1051/0004-6361/201525821
[arXiv:1502.01594 [astro-ph.CO]].
%248 citations counted in INSPIRE as of 10 Dec 2020


%\cite{Chowdhury:2018mhj}
\bibitem{Chowdhury:2018mhj}
D.~Chowdhury, L.~Sriramkumar and M.~Kamionkowski,
%``Enhancing the cross-correlations between magnetic fields and scalar perturbations through parity violation,''
JCAP \textbf{10} (2018), 031
%doi:10.1088/1475-7516/2018/10/031
[arXiv:1807.07477 [astro-ph.CO]].
%6 citations counted in INSPIRE as of 10 Dec 2020


%\cite{Vachaspati:1991nm}
\bibitem{Vachaspati:1991nm}
T.~Vachaspati,
%``Magnetic fields from cosmological phase transitions,''
Phys. Lett. B \textbf{265} (1991), 258-261
%doi:10.1016/0370-2693(91)90051-Q
%546 citations counted in INSPIRE as of 10 Dec 2020


%\cite{Turner:1987bw}
\bibitem{Turner:1987bw}
M.~S.~Turner and L.~M.~Widrow,
%``Inflation Produced, Large Scale Magnetic Fields,''
Phys. Rev. D \textbf{37} (1988), 2743
%doi:10.1103/PhysRevD.37.2743
%781 citations counted in INSPIRE as of 10 Dec 2020


%\cite{Takahashi:2005nd}
\bibitem{Takahashi:2005nd}
K.~Takahashi, K.~Ichiki, H.~Ohno and H.~Hanayama,
%``Magnetic field generation from cosmological perturbations,''
Phys. Rev. Lett. \textbf{95} (2005), 121301
%doi:10.1103/PhysRevLett.95.121301
[arXiv:astro-ph/0502283 [astro-ph]].
%107 citations counted in INSPIRE as of 10 Dec 2020

%\cite{Agullo:2013tba}
\bibitem{Agullo:2013tba}
I.~Agullo and J.~Navarro-Salas,
%``Conformal anomaly and primordial magnetic fields,''
[arXiv:1309.3435 [gr-qc]].
%13 citations counted in INSPIRE as of 10 Dec 2020


%\cite{Ferreira:2013sqa}
\bibitem{Ferreira:2013sqa}
R.~J.~Z.~Ferreira, R.~K.~Jain and M.~S.~Sloth,
%``Inflationary magnetogenesis  without the strong coupling problem,''
JCAP \textbf{10} (2013), 004
%doi:10.1088/1475-7516/2013/10/004
[arXiv:1305.7151 [astro-ph.CO]].
%94 citations counted in INSPIRE as of 10 Dec 2020

%\cite{Atmjeet:2014cxa}
\bibitem{Atmjeet:2014cxa}
K.~Atmjeet, T.~R.~Seshadri and K.~Subramanian,
%``Helical cosmological magnetic fields from extra-dimensions,''
Phys. Rev. D \textbf{91} (2015), 103006
%doi:10.1103/PhysRevD.91.103006
[arXiv:1409.6840 [astro-ph.CO]].
%12 citations counted in INSPIRE as of 10 Dec 2020

%\cite{Frion:2020bxc}
\bibitem{Frion:2020bxc}
E.~Frion, N.~Pinto-Neto, S.~D.~P.~Vitenti and S.~E.~Perez Bergliaffa,
%``Primordial Magnetogenesis in a Bouncing Universe,''
Phys. Rev. D \textbf{101} (2020) no.10, 103503
%doi:10.1103/PhysRevD.101.103503
[arXiv:2004.07269 [gr-qc]].
%3 citations counted in INSPIRE as of 10 Dec 2020

%\cite{Chowdhury:2016aet}
\bibitem{Chowdhury:2016aet}
D.~Chowdhury, L.~Sriramkumar and R.~K.~Jain,
%``Duality and scale invariant magnetic fields from bouncing universes,''
Phys. Rev. D \textbf{94} (2016) no.8, 083512
%doi:10.1103/PhysRevD.94.083512
[arXiv:1604.02143 [gr-qc]].
%18 citations counted in INSPIRE as of 10 Dec 2020


%\cite{Chowdhury:2018blx}
\bibitem{Chowdhury:2018blx}
D.~Chowdhury, L.~Sriramkumar and M.~Kamionkowski,
%``Cross-correlations between scalar perturbations and magnetic fields in bouncing universes,''
JCAP \textbf{01} (2019), 048
%doi:10.1088/1475-7516/2019/01/048
[arXiv:1807.05530 [astro-ph.CO]].
%12 citations counted in INSPIRE as of 10 Dec 2020

%\cite{Qian:2016lbf}
\bibitem{Qian:2016lbf}
P.~Qian, Y.~F.~Cai, D.~A.~Easson and Z.~K.~Guo,
%``Magnetogenesis in bouncing cosmology,''
Phys. Rev. D \textbf{94} (2016) no.8, 083524
%doi:10.1103/PhysRevD.94.083524
[arXiv:1607.06578 [gr-qc]].
%13 citations counted in INSPIRE as of 10 Dec 2020

%\cite{Koley:2016jdw}
\bibitem{Koley:2016jdw}
R.~Koley and S.~Samtani,
%``Magnetogenesis in Matter - Ekpyrotic Bouncing Cosmology,''
JCAP \textbf{04} (2017), 030
%doi:10.1088/1475-7516/2017/04/030
[arXiv:1612.08556 [gr-qc]].
%10 citations counted in INSPIRE as of 10 Dec 2020


%\cite{Membiela:2013cea}
\bibitem{Membiela:2013cea}
F.~A.~Membiela,
%``Primordial magnetic fields from a non-singular bouncing cosmology,''
Nucl. Phys. B \textbf{885} (2014), 196-224
%doi:10.1016/j.nuclphysb.2014.05.018
[arXiv:1312.2162 [astro-ph.CO]].
%19 citations counted in INSPIRE as of 10 Dec 2020


 \bibitem{guth}
A.H. Guth;  Phys.Rev. D23 347-356 (1981).

  %\cite{Linde:2005ht}
\bibitem{Linde:2005ht}
  A.~D.~Linde,
  %``Particle physics and inflationary cosmology,''
  Contemp.\ Concepts Phys.\  {\bf 5} (1990) 1
  [hep-th/0503203].
  %%CITATION = HEP-TH/0503203;%%
  %676 citations counted in INSPIRE as of 21 Oct 2019


 %\cite{Langlois:2004de}
\bibitem{Langlois:2004de}
  D.~Langlois,
  %``Inflation, quantum fluctuations and cosmological perturbations,''
  hep-th/0405053.
  %%CITATION = HEP-TH/0405053;%%
  %61 citations counted in INSPIRE as of 21 Oct 2019


  %\cite{Riotto:2002yw}
\bibitem{Riotto:2002yw}
  A.~Riotto,
  %``Inflation and the theory of cosmological perturbations,''
  ICTP Lect.\ Notes Ser.\  {\bf 14} (2003) 317
  [hep-ph/0210162].
  %%CITATION = HEP-PH/0210162;%%
  %313 citations counted in INSPIRE as of 21 Oct 2019




%\cite{Baumann:2009ds}
\bibitem{Baumann:2009ds}
D.~Baumann,
%``Inflation,''
%doi:10.1142/9789814327183 0010
[arXiv:0907.5424 [hep-th]].
%729 citations counted in INSPIRE as of 24 Aug 2020


%\cite{Bamba:2015uma}
\bibitem{Bamba:2015uma}
K.~Bamba and S.~D.~Odintsov,
%``Inflationary cosmology in modified gravity theories,''
Symmetry \textbf{7} (2015) no.1, 220-240
%doi:10.3390/sym7010220
[arXiv:1503.00442 [hep-th]].
%205 citations counted in INSPIRE as of 11 Dec 2020




%\cite{Caprini:2014mja}
\bibitem{Caprini:2014mja}
C.~Caprini and L.~Sorbo,
%``Adding helicity to inflationary magnetogenesis,''
JCAP \textbf{10} (2014), 056
%doi:10.1088/1475-7516/2014/10/056
[arXiv:1407.2809 [astro-ph.CO]].
%108 citations counted in INSPIRE as of 10 Dec 2020


%\cite{Kobayashi:2014sga}
\bibitem{Kobayashi:2014sga}
T.~Kobayashi,
%``Primordial Magnetic Fields from the Post-Inflationary Universe,''
JCAP \textbf{05} (2014), 040
%doi:10.1088/1475-7516/2014/05/040
[arXiv:1403.5168 [astro-ph.CO]].
%45 citations counted in INSPIRE as of 10 Dec 2020


%\cite{Atmjeet:2013yta}
\bibitem{Atmjeet:2013yta}
K.~Atmjeet, I.~Pahwa, T.~R.~Seshadri and K.~Subramanian,
%``Cosmological Magnetogenesis From Extra-dimensional Gauss Bonnet Gravity,''
Phys. Rev. D \textbf{89} (2014) no.6, 063002
%doi:10.1103/PhysRevD.89.063002
[arXiv:1312.5815 [astro-ph.CO]].
%18 citations counted in INSPIRE as of 10 Dec 2020


%\cite{Fujita:2015iga}
\bibitem{Fujita:2015iga}
T.~Fujita, R.~Namba, Y.~Tada, N.~Takeda and H.~Tashiro,
%``Consistent generation of magnetic fields in axion inflation models,''
JCAP \textbf{05} (2015), 054
%doi:10.1088/1475-7516/2015/05/054
[arXiv:1503.05802 [astro-ph.CO]].
%70 citations counted in INSPIRE as of 10 Dec 2020


%\cite{Campanelli:2015jfa}
\bibitem{Campanelli:2015jfa}
L.~Campanelli,
%``Lorentz-violating inflationary magnetogenesis,''
Eur. Phys. J. C \textbf{75} (2015) no.6, 278
%doi:10.1140/epjc/s10052-015-3510-x
[arXiv:1503.07415 [gr-qc]].
%23 citations counted in INSPIRE as of 10 Dec 2020


%\cite{Tasinato:2014fia}
\bibitem{Tasinato:2014fia}
G.~Tasinato,
%``A scenario for inflationary magnetogenesis without strong coupling problem,''
JCAP \textbf{03} (2015), 040
%doi:10.1088/1475-7516/2015/03/040
[arXiv:1411.2803 [hep-th]].
%31 citations counted in INSPIRE as of 10 Dec 2020


%\cite{Campanelli:2007cg}
\bibitem{Campanelli:2007cg}
L.~Campanelli, P.~Cea, G.~L.~Fogli and L.~Tedesco,
%``Inflation-Produced Magnetic Fields in Nonlinear Electrodynamics,''
Phys. Rev. D \textbf{77} (2008), 043001
%doi:10.1103/PhysRevD.77.043001
[arXiv:0710.2993 [astro-ph]].
%54 citations counted in INSPIRE as of 10 Dec 2020




%\cite{Urban:2013aka}
\bibitem{Urban:2013aka}
F.~R.~Urban,
%``The anisotropy of a three- and a one-form,''
JCAP \textbf{08} (2013), 008
%doi:10.1088/1475-7516/2013/08/008
[arXiv:1306.6429 [astro-ph.CO]].
%5 citations counted in INSPIRE as of 10 Dec 2020


%\cite{Markkanen:2017kmy}
\bibitem{Markkanen:2017kmy}
T.~Markkanen, S.~Nurmi, S.~Rasanen and V.~Vennin,
%``Narrowing the window of inflationary magnetogenesis,''
JCAP \textbf{06} (2017), 035
%doi:10.1088/1475-7516/2017/06/035
[arXiv:1704.01343 [astro-ph.CO]].
%5 citations counted in INSPIRE as of 10 Dec 2020


%\cite{Brandenberger:2012zb}
\bibitem{Brandenberger:2012zb}
R.~H.~Brandenberger,
%``The Matter Bounce Alternative to Inflationary Cosmology,''
arXiv:1206.4196 [astro-ph.CO].
%%CITATION = ARXIV:1206.4196;%%
%110 citations counted in INSPIRE as of 29 Apr 2017



%\cite{Brandenberger:2016vhg}
\bibitem{Brandenberger:2016vhg}
R.~Brandenberger and P.~Peter,
%``Bouncing Cosmologies: Progress and Problems,''
arXiv:1603.05834 [hep-th].
%%CITATION = ARXIV:1603.05834;%%
%40 citations counted in INSPIRE as of 04 Feb 2017


%\cite{Battefeld:2014uga}
\bibitem{Battefeld:2014uga}
 D.~Battefeld and P.~Peter,
 %``A Critical Review of Classical Bouncing Cosmologies,''
 Phys.\ Rept.\ {\bf 571} (2015) 1
 %doi:10.1016/j.physrep.2014.12.004
 [arXiv:1406.2790 [astro-ph.CO]].
 %%CITATION = doi:10.1016/j.physrep.2014.12.004;%%
 %75 citations counted in INSPIRE as of 23 Dec 2016


%\cite{Novello:2008ra}
\bibitem{Novello:2008ra}
M.~Novello and S.~E.~P.~Bergliaffa,
 ``Bouncing Cosmologies,''
Phys.\ Rept.\ {\bf 463} (2008) 127
%doi:10.1016/j.physrep.2008.04.006
[arXiv:0802.1634 [astro-ph]].
%%CITATION = doi:10.1016/j.physrep.2008.04.006;%%
%306 citations counted in INSPIRE as of 24 Sep 2016


%\cite{Cai:2014bea}
\bibitem{Cai:2014bea}
Y.~F.~Cai,
%``Exploring Bouncing Cosmologies with Cosmological Surveys,''
Sci.\ China Phys.\ Mech.\ Astron.\  {\bf 57} (2014) 1414
%doi:10.1007/s11433-014-5512-3
[arXiv:1405.1369 [hep-th]].
%%CITATION = doi:10.1007/s11433-014-5512-3;%%
%52 citations counted in INSPIRE as of 29 Apr 2017


%\cite{Nojiri:2019lqw}
\bibitem{Nojiri:2019lqw}
S.~Nojiri, S.~D.~Odintsov, V.~K.~Oikonomou and T.~Paul,
%``Nonsingular bounce cosmology from Lagrange multiplier $F(R)$ gravity,''
Phys. Rev. D \textbf{100} (2019) no.8, 084056
%doi:10.1103/PhysRevD.100.084056
[arXiv:1910.03546 [gr-qc]].
%7 citations counted in INSPIRE as of 10 Dec 2020


%\cite{Odintsov:2015ynk}
\bibitem{Odintsov:2015ynk}
S.~D.~Odintsov and V.~K.~Oikonomou,
%``Big-Bounce with Finite-time Singularity: The $F(R)$ Gravity Description,''
Int. J. Mod. Phys. D \textbf{26} (2017) no.08, 1750085
%doi:10.1142/S0218271817500857
[arXiv:1512.04787 [gr-qc]].
%54 citations counted in INSPIRE as of 10 Dec 2020


%\cite{Cai:2008qw}
\bibitem{Cai:2008qw}
Y.~F.~Cai, T.~t.~Qiu, R.~Brandenberger and X.~m.~Zhang,
%``A Nonsingular Cosmology with a Scale-Invariant Spectrum of Cosmological Perturbations from Lee-Wick Theory,''
Phys. Rev. D \textbf{80} (2009), 023511
%doi:10.1103/PhysRevD.80.023511
[arXiv:0810.4677 [hep-th]].
%181 citations counted in INSPIRE as of 10 Dec 2020


%\cite{Cai:2016thi}
\bibitem{Cai:2016thi}
Y.~Cai, Y.~Wan, H.~G.~Li, T.~Qiu and Y.~S.~Piao,
%``The Effective Field Theory of nonsingular cosmology,''
JHEP \textbf{01} (2017), 090
%doi:10.1007/JHEP01(2017)090
[arXiv:1610.03400 [gr-qc]].
%89 citations counted in INSPIRE as of 27 Nov 2020


%\cite{Elizalde:2019tee}
\bibitem{Elizalde:2019tee}
E.~Elizalde, S.~D.~Odintsov and T.~Paul,
%``Viable non-singular cosmic bounce in holonomy improved F(R) gravity endowed with a Lagrange multiplier,''
Eur. Phys. J. C \textbf{80} (2020) no.1, 10
%doi:10.1140/epjc/s10052-019-7544-3
[arXiv:1912.05138 [gr-qc]].
%4 citations counted in INSPIRE as of 10 Dec 2020


%\cite{Elizalde:2020zcb}
\bibitem{Elizalde:2020zcb}
E.~Elizalde, S.~D.~Odintsov, V.~K.~Oikonomou and T.~Paul,
%``Extended matter bounce scenario in ghost free $f(R,\mathcal{G})$ gravity compatible with GW170817,''
Nucl. Phys. B \textbf{954} (2020), 114984
%doi:10.1016/j.nuclphysb.2020.114984
[arXiv:2003.04264 [gr-qc]].
%15 citations counted in INSPIRE as of 10 Dec 2020


%\cite{Navo:2020eqt}
\bibitem{Navo:2020eqt}
G.~Nav\'o and E.~Elizalde,
%``Stability of hyperbolic and matter-dominated bounce cosmologies from $F(R,\mathcal G$)modified gravity at late evolution stages,''
Int. J. Geom. Meth. Mod. Phys. \textbf{17} (2020) no.11, 2050162
%doi:10.1142/S0219887820501625
[arXiv:2007.11507 [gr-qc]].
%1 citations counted in INSPIRE as of 11 Dec 2020


%\cite{Bamba:2014mya}
\bibitem{Bamba:2014mya}
K.~Bamba, A.~N.~Makarenko, A.~N.~Myagky and S.~D.~Odintsov,
%``Bouncing cosmology in modified Gauss-Bonnet gravity,''
Phys. Lett. B \textbf{732} (2014), 349-355
%doi:10.1016/j.physletb.2014.04.004
[arXiv:1403.3242 [hep-th]].
%78 citations counted in INSPIRE as of 11 Dec 2020


%\cite{Odintsov:2020zct}
\bibitem{Odintsov:2020zct}
S.~D.~Odintsov, V.~K.~Oikonomou and T.~Paul,
%``From a Bounce to the Dark Energy Era with $F(R)$ Gravity,''
Class. Quant. Grav. \textbf{37} (2020) no.23, 235005
%doi:10.1088/1361-6382/abbc47
[arXiv:2009.09947 [gr-qc]].
%1 citations counted in INSPIRE as of 10 Dec 2020


%\cite{Banerjee:2020uil}
\bibitem{Banerjee:2020uil}
I.~Banerjee, T.~Paul and S.~SenGupta,
%``Bouncing cosmology in a curved braneworld,''
[arXiv:2011.11886 [gr-qc]].
%0 citations counted in INSPIRE as of 10 Dec 2020











%\cite{Li:2007jm}
\bibitem{Li:2007jm}
B.~Li, J.~D.~Barrow and D.~F.~Mota,
%``The Cosmology of Modified Gauss-Bonnet Gravity,''
Phys. Rev. D \textbf{76} (2007), 044027
%doi:10.1103/PhysRevD.76.044027
[arXiv:0705.3795 [gr-qc]].
%197 citations counted in INSPIRE as of 02 May 2020

%\cite{Odintsov:2018nch}
\bibitem{Odintsov:2018nch}
S.~D.~Odintsov, V.~K.~Oikonomou and S.~Banerjee,
%``Dynamics of inflation and dark energy from $F(R,G)$ gravity,''
Nucl.\ Phys.\ B {\bf 938} (2019) 935
%doi:10.1016/j.nuclphysb.2018.07.013
[arXiv:1807.00335 [gr-qc]].
%%CITATION = doi:10.1016/j.nuclphysb.2018.07.013;%%
%16 citations counted in INSPIRE as of 02 May 2020

%\cite{Carter:2005fu}
\bibitem{Carter:2005fu}
B.~M.~Carter and I.~P.~Neupane,
%``Towards inflation and dark energy cosmologies from modified Gauss-Bonnet theory,''
JCAP \textbf{06} (2006), 004
%doi:10.1088/1475-7516/2006/06/004
[arXiv:hep-th/0512262 [hep-th]].
%131 citations counted in INSPIRE as of 02 May 2020



%\cite{Nojiri:2019dwl}
\bibitem{Nojiri:2019dwl}
S.~Nojiri, S.~Odintsov, V.~Oikonomou, N.~Chatzarakis and T.~Paul,
%``Viable inflationary models in a ghost-free Gauss–Bonnet theory of gravity,''
Eur. Phys. J. C \textbf{79} (2019) no.7, 565
%doi:10.1140/epjc/s10052-019-7080-1
[arXiv:1907.00403 [gr-qc]].
%6 citations counted in INSPIRE as of 12 Jun 2020


%\cite{Elizalde:2010jx}
\bibitem{Elizalde:2010jx}
E.~Elizalde, R.~Myrzakulov, V.~Obukhov and D.~Saez-Gomez,
%``LambdaCDM epoch reconstruction from F(R,G) and modified Gauss-Bonnet gravities,''
Class. Quant. Grav. \textbf{27} (2010), 095007
%doi:10.1088/0264-9381/27/9/095007
[arXiv:1001.3636 [gr-qc]].
%132 citations counted in INSPIRE as of 02 May 2020

%\cite{Makarenko:2016jsy}
\bibitem{Makarenko:2016jsy}
A.~N.~Makarenko,
%``The role of Lagrange multiplier in Gauss–Bonnet dark energy,''
Int. J. Geom. Meth. Mod. Phys. \textbf{13} (2016) no.05, 1630006
%doi:10.1142/S0219887816300063
%5 citations counted in INSPIRE as of 02 May 2020

%\cite{delaCruzDombriz:2011wn}
\bibitem{delaCruzDombriz:2011wn}
A.~de la Cruz-Dombriz and D.~Saez-Gomez,
%``On the stability of the cosmological solutions in $f(R,G)$ gravity,''
Class. Quant. Grav. \textbf{29} (2012), 245014
%doi:10.1088/0264-9381/29/24/245014
[arXiv:1112.4481 [gr-qc]].
%61 citations counted in INSPIRE as of 02 May 2020


%\cite{Bamba:2007ef}
\bibitem{Bamba:2007ef}
K.~Bamba, Z.~K.~Guo and N.~Ohta,
%``Accelerating Cosmologies in the Einstein-Gauss-Bonnet Theory with Dilaton,''
Prog. Theor. Phys. \textbf{118} (2007), 879-892
%doi:10.1143/PTP.118.879
[arXiv:0707.4334 [hep-th]].
%86 citations counted in INSPIRE as of 11 Dec 2020



%\cite{Chakraborty:2018scm}
\bibitem{Chakraborty:2018scm}
S.~Chakraborty, T.~Paul and S.~SenGupta,
%``Inflation driven by Einstein-Gauss-Bonnet gravity,''
Phys. Rev. D \textbf{98} (2018) no.8, 083539
%doi:10.1103/PhysRevD.98.083539
[arXiv:1804.03004 [gr-qc]].
%25 citations counted in INSPIRE as of 02 May 2020

%\cite{Kanti:2015pda}
\bibitem{Kanti:2015pda}
P.~Kanti, R.~Gannouji and N.~Dadhich,
%``Gauss-Bonnet Inflation,''
Phys. Rev. D \textbf{92} (2015) no.4, 041302
%doi:10.1103/PhysRevD.92.041302
[arXiv:1503.01579 [hep-th]].
%60 citations counted in INSPIRE as of 02 May 2020

%\cite{Kanti:2015dra}
\bibitem{Kanti:2015dra}
P.~Kanti, R.~Gannouji and N.~Dadhich,
%``Early-time cosmological solutions in Einstein-scalar-Gauss-Bonnet theory,''
Phys. Rev. D \textbf{92} (2015) no.8, 083524
%doi:10.1103/PhysRevD.92.083524
[arXiv:1506.04667 [hep-th]].
%31 citations counted in INSPIRE as of 02 May 2020

%\cite{Odintsov:2018zhw}
\bibitem{Odintsov:2018zhw}
S.~D.~Odintsov and V.~K.~Oikonomou,
%``Viable Inflation in Scalar-Gauss-Bonnet Gravity and Reconstruction from Observational Indices,''
Phys.\ Rev.\ D {\bf 98} (2018) no.4,  044039
%doi:10.1103/PhysRevD.98.044039
[arXiv:1808.05045 [gr-qc]].
%%CITATION = doi:10.1103/PhysRevD.98.044039;%%
%24 citations counted in INSPIRE as of 02 May 2020

%\cite{Saridakis:2017rdo}
\bibitem{Saridakis:2017rdo}
E.~N.~Saridakis,
%``Ricci-Gauss-Bonnet holographic dark energy,''
Phys.\ Rev.\ D {\bf 97} (2018) no.6,  064035
%doi:10.1103/PhysRevD.97.064035
[arXiv:1707.09331 [gr-qc]].
%%CITATION = doi:10.1103/PhysRevD.97.064035;%%
%13 citations counted in INSPIRE as of 16 Nov 2019

%\cite{Cognola:2006eg}
\bibitem{Cognola:2006eg}
G.~Cognola, E.~Elizalde, S.~Nojiri, S.~D.~Odintsov and S.~Zerbini,
%``Dark energy in modified Gauss-Bonnet gravity: Late-time acceleration and the hierarchy problem,''
Phys. Rev. D \textbf{73} (2006), 084007
%doi:10.1103/PhysRevD.73.084007
[arXiv:hep-th/0601008 [hep-th]].
%506 citations counted in INSPIRE as of 02 May 2020


%\cite{Dai:2014jja}
\bibitem{Dai:2014jja}
L.~Dai, M.~Kamionkowski and J.~Wang,
%``Reheating constraints to inflationary models,''
Phys. Rev. Lett. \textbf{113} (2014), 041302
%doi:10.1103/PhysRevLett.113.041302
[arXiv:1404.6704 [astro-ph.CO]].
%149 citations counted in INSPIRE as of 10 Dec 2020


%\cite{Albrecht:1982mp}
\bibitem{Albrecht:1982mp}
A.~Albrecht, P.~J.~Steinhardt, M.~S.~Turner and F.~Wilczek,
%``Reheating an Inflationary Universe,''
Phys. Rev. Lett. \textbf{48} (1982), 1437
%doi:10.1103/PhysRevLett.48.1437
%497 citations counted in INSPIRE as of 10 Dec 2020


%\cite{Ellis:2015pla}
\bibitem{Ellis:2015pla}
J.~Ellis, M.~A.~G.~Garcia, D.~V.~Nanopoulos and K.~A.~Olive,
%``Calculations of Inflaton Decays and Reheating: with Applications to No-Scale Inflation Models,''
JCAP \textbf{07} (2015), 050
%doi:10.1088/1475-7516/2015/07/050
[arXiv:1505.06986 [hep-ph]].
%58 citations counted in INSPIRE as of 10 Dec 2020


%\cite{Ueno:2016dim}
\bibitem{Ueno:2016dim}
Y.~Ueno and K.~Yamamoto,
%``Constraints on $\alpha$-attractor inflation and reheating,''
Phys. Rev. D \textbf{93} (2016) no.8, 083524
%doi:10.1103/PhysRevD.93.083524
[arXiv:1602.07427 [astro-ph.CO]].
%47 citations counted in INSPIRE as of 10 Dec 2020


%\cite{Eshaghi:2016kne}
\bibitem{Eshaghi:2016kne}
M.~Eshaghi, M.~Zarei, N.~Riazi and A.~Kiasatpour,
%``CMB and reheating constraints to $\alpha$-attractor inflationary models,''
Phys. Rev. D \textbf{93} (2016) no.12, 123517
%doi:10.1103/PhysRevD.93.123517
[arXiv:1602.07914 [astro-ph.CO]].
%28 citations counted in INSPIRE as of 10 Dec 2020


%\cite{Maity:2018qhi}
\bibitem{Maity:2018qhi}
D.~Maity and P.~Saha,
%``(P)reheating after minimal Plateau Inflation and constraints from CMB,''
JCAP \textbf{07} (2019), 018
%doi:10.1088/1475-7516/2019/07/018
[arXiv:1811.11173 [astro-ph.CO]].
%8 citations counted in INSPIRE as of 10 Dec 2020


%\cite{Haque:2020zco}
\bibitem{Haque:2020zco}
M.~R.~Haque, D.~Maity and P.~Saha,
%``Two-phase reheating: CMB constraints on inflation and dark matter phenomenology,''
Phys. Rev. D \textbf{102} (2020) no.8, 083534
%doi:10.1103/PhysRevD.102.083534
[arXiv:2009.02794 [hep-th]].
%0 citations counted in INSPIRE as of 10 Dec 2020


%\cite{DiMarco:2017zek}
\bibitem{DiMarco:2017zek}
A.~Di Marco, P.~Cabella and N.~Vittorio,
%``Constraining the general reheating phase in the $\alpha$-attractor inflationary cosmology,''
Phys. Rev. D \textbf{95} (2017) no.10, 103502
%doi:10.1103/PhysRevD.95.103502
[arXiv:1705.04622 [astro-ph.CO]].
%9 citations counted in INSPIRE as of 10 Dec 2020


%\cite{Drewes:2017fmn}
\bibitem{Drewes:2017fmn}
M.~Drewes, J.~U.~Kang and U.~R.~Mun,
%``CMB constraints on the inflaton couplings and reheating temperature in $\alpha$-attractor inflation,''
JHEP \textbf{11} (2017), 072
%doi:10.1007/JHEP11(2017)072
[arXiv:1708.01197 [astro-ph.CO]].
%15 citations counted in INSPIRE as of 10 Dec 2020


%\cite{DiMarco:2018bnw}
\bibitem{DiMarco:2018bnw}
A.~Di Marco, G.~Pradisi and P.~Cabella,
%``Inflationary scale, reheating scale, and pre-BBN cosmology with scalar fields,''
Phys. Rev. D \textbf{98} (2018) no.12, 123511
%doi:10.1103/PhysRevD.98.123511
[arXiv:1807.05916 [astro-ph.CO]].
%17 citations counted in INSPIRE as of 10 Dec 2020





%\cite{Kushwaha:2020nfa}
\bibitem{Kushwaha:2020nfa}
A.~Kushwaha and S.~Shankaranarayanan,
%``Helical magnetic fields from Riemann coupling,''
Phys. Rev. D \textbf{102} (2020) no.10, 103528
%doi:10.1103/PhysRevD.102.103528
[arXiv:2008.10825 [gr-qc]].
%0 citations counted in INSPIRE as of 10 Dec 2020


%\cite{Guo:2015awg}
\bibitem{Guo:2015awg}
P.~Qian and Z.~K.~Guo,
%``Model of inflationary magnetogenesis,''
Phys. Rev. D \textbf{93} (2016) no.4, 043541
%doi:10.1103/PhysRevD.93.043541
[arXiv:1512.05050 [astro-ph.CO]].
%6 citations counted in INSPIRE as of 10 Dec 2020


%\cite{Haque:2020bip}
\bibitem{Haque:2020bip}
M.~R.~Haque, D.~Maity and S.~Pal,
%``Probing the reheating phase through primordial magnetic field and CMB,''
[arXiv:2012.10859 [hep-th]].
%0 citations counted in INSPIRE as of 23 Dec 2020


%\cite{Odintsov:2019clh}
\bibitem{Odintsov:2019clh}
S.~D.~Odintsov and V.~K.~Oikonomou,
%``Inflationary Phenomenology of Einstein Gauss-Bonnet Gravity Compatible with GW170817,''
Phys. Lett. B \textbf{797} (2019), 134874
%doi:10.1016/j.physletb.2019.134874
[arXiv:1908.07555 [gr-qc]].
%18 citations counted in INSPIRE as of 10 Dec 2020


%\cite{Odintsov:2020sqy}
\bibitem{Odintsov:2020sqy}
S.~D.~Odintsov, V.~K.~Oikonomou and F.~P.~Fronimos,
%``Rectifying Einstein-Gauss-Bonnet Inflation in View of GW170817,''
Nucl. Phys. B \textbf{958} (2020), 115135
%doi:10.1016/j.nuclphysb.2020.115135
[arXiv:2003.13724 [gr-qc]].
%27 citations counted in INSPIRE as of 10 Dec 2020


%\cite{Odintsov:2020mkz}
\bibitem{Odintsov:2020mkz}
S.~D.~Odintsov, V.~K.~Oikonomou, F.~P.~Fronimos and S.~A.~Venikoudis,
%``GW170817-compatible constant-roll Einstein\textendash{}Gauss\textendash{}Bonnet inflation and non-Gaussianities,''
Phys. Dark Univ. \textbf{30} (2020), 100718
%doi:10.1016/j.dark.2020.100718
[arXiv:2009.06113 [gr-qc]].
%3 citations counted in INSPIRE as of 10 Dec 2020





%\cite{Cook:2015vqa}
\bibitem{Cook:2015vqa}
J.~L.~Cook, E.~Dimastrogiovanni, D.~A.~Easson and L.~M.~Krauss,
%``Reheating predictions in single field inflation,''
JCAP \textbf{04} (2015), 047
%doi:10.1088/1475-7516/2015/04/047
[arXiv:1502.04673 [astro-ph.CO]].
%108 citations counted in INSPIRE as of 10 Dec 2020


%\cite{Kobayashi:2014zza}
\bibitem{Kobayashi:2014zza}
T.~Kobayashi and N.~Afshordi,
%``Schwinger Effect in 4D de Sitter Space and Constraints on Magnetogenesis in the Early Universe,''
JHEP \textbf{10} (2014), 166
doi:10.1007/JHEP10(2014)166
[arXiv:1408.4141 [hep-th]].
%90 citations counted in INSPIRE as of 08 Feb 2021


%\cite{Stahl:2018idd}
\bibitem{Stahl:2018idd}
C.~Stahl,
%``Schwinger effect impacting primordial magnetogenesis,''
Nucl. Phys. B \textbf{939} (2019), 95-104
doi:10.1016/j.nuclphysb.2018.12.017
[arXiv:1806.06692 [hep-th]].
%17 citations counted in INSPIRE as of 08 Feb 2021


%\cite{Rajeev:2019okd}
\bibitem{Rajeev:2019okd}
K.~Rajeev, S.~Chakraborty and T.~Padmanabhan,
%``Generalized Schwinger effect and particle production in an expanding universe,''
Phys. Rev. D \textbf{100} (2019) no.4, 045019
doi:10.1103/PhysRevD.100.045019
[arXiv:1904.03207 [gr-qc]].
%3 citations counted in INSPIRE as of 08 Feb 2021


%\cite{Fujita:2013qxa}
\bibitem{Fujita:2013qxa}
T.~Fujita and S.~Yokoyama,
%``Higher order statistics of curvature perturbations in IFF model and its Planck constraints,''
JCAP \textbf{09} (2013), 009
doi:10.1088/1475-7516/2013/09/009
[arXiv:1306.2992 [astro-ph.CO]].
%60 citations counted in INSPIRE as of 08 Feb 2021


%\cite{Fujita:2016qab}
\bibitem{Fujita:2016qab}
T.~Fujita and R.~Namba,
%``Pre-reheating Magnetogenesis in the Kinetic Coupling Model,''
Phys. Rev. D \textbf{94} (2016) no.4, 043523
doi:10.1103/PhysRevD.94.043523
[arXiv:1602.05673 [astro-ph.CO]].
%35 citations counted in INSPIRE as of 08 Feb 2021


%\cite{Barnaby:2012tk}
\bibitem{Barnaby:2012tk}
N.~Barnaby, R.~Namba and M.~Peloso,
%``Observable non-gaussianity from gauge field production in slow roll inflation, and a challenging connection with magnetogenesis,''
Phys. Rev. D \textbf{85} (2012), 123523
doi:10.1103/PhysRevD.85.123523
[arXiv:1202.1469 [astro-ph.CO]].
%142 citations counted in INSPIRE as of 08 Feb 2021


%\cite{Ferreira:2014hma}
\bibitem{Ferreira:2014hma}
R.~J.~Z.~Ferreira, R.~K.~Jain and M.~S.~Sloth,
%``Inflationary Magnetogenesis without the Strong Coupling Problem II: Constraints from CMB anisotropies and B-modes,''
JCAP \textbf{06} (2014), 053
doi:10.1088/1475-7516/2014/06/053
[arXiv:1403.5516 [astro-ph.CO]].
%66 citations counted in INSPIRE as of 08 Feb 2021


%\cite{Giovannini:2013rme}
\bibitem{Giovannini:2013rme}
M.~Giovannini,
%``Fluctuations of inflationary magnetogenesis,''
Phys. Rev. D \textbf{87} (2013) no.8, 083004
doi:10.1103/PhysRevD.87.083004
[arXiv:1302.2243 [hep-th]].
%19 citations counted in INSPIRE as of 08 Feb 2021


%\cite{Bamba:2014vda}
\bibitem{Bamba:2014vda}
K.~Bamba,
%``Generation of large-scale magnetic fields, non-Gaussianity, and primordial gravitational waves in inflationary cosmology,''
Phys. Rev. D \textbf{91} (2015), 043509
doi:10.1103/PhysRevD.91.043509
[arXiv:1411.4335 [astro-ph.CO]].
%20 citations counted in INSPIRE as of 08 Feb 2021




%\cite{Suyama:2012wh}
\bibitem{Suyama:2012wh}
T.~Suyama and J.~Yokoyama,
%``Metric perturbation from inflationary magnetic field and generic bound on inflation models,''
Phys. Rev. D \textbf{86} (2012), 023512
doi:10.1103/PhysRevD.86.023512
[arXiv:1204.3976 [astro-ph.CO]].
%38 citations counted in INSPIRE as of 08 Feb 2021


%\cite{Himmetoglu:2009qi}
\bibitem{Himmetoglu:2009qi}
B.~Himmetoglu, C.~R.~Contaldi and M.~Peloso,
%``Ghost instabilities of cosmological models with vector fields nonminimally coupled to the curvature,''
Phys. Rev. D \textbf{80} (2009), 123530
doi:10.1103/PhysRevD.80.123530
[arXiv:0909.3524 [astro-ph.CO]].
%123 citations counted in INSPIRE as of 08 Feb 2021


%\cite{Himmetoglu:2008zp}
\bibitem{Himmetoglu:2008zp}
B.~Himmetoglu, C.~R.~Contaldi and M.~Peloso,
%``Instability of anisotropic cosmological solutions supported by vector fields,''
Phys. Rev. Lett. \textbf{102} (2009), 111301
doi:10.1103/PhysRevLett.102.111301
[arXiv:0809.2779 [astro-ph]].
%186 citations counted in INSPIRE as of 08 Feb 2021


%\cite{Himmetoglu:2008hx}
\bibitem{Himmetoglu:2008hx}
B.~Himmetoglu, C.~R.~Contaldi and M.~Peloso,
%``Instability of the ACW model, and problems with massive vectors during inflation,''
Phys. Rev. D \textbf{79} (2009), 063517
doi:10.1103/PhysRevD.79.063517
[arXiv:0812.1231 [astro-ph]].
%128 citations counted in INSPIRE as of 08 Feb 2021


%\cite{Karciauskas:2010as}
\bibitem{Karciauskas:2010as}
M.~Karciauskas and D.~H.~Lyth,
%``On the health of a vector field with (R A\textasciicircum{}2)/6 coupling to gravity,''
JCAP \textbf{11} (2010), 023
doi:10.1088/1475-7516/2010/11/023
[arXiv:1007.1426 [astro-ph.CO]].
%27 citations counted in INSPIRE as of 08 Feb 2021

\end{thebibliography}
\end{document}